\documentclass[conference]{IEEEtran}
\usepackage{cite}
\usepackage{amsmath,amssymb,amsfonts}
\usepackage{algorithmic}
\usepackage{graphicx}
\usepackage{textcomp}
\usepackage{xcolor}
\usepackage{fancyhdr}

\usepackage{mathptmx} 
\usepackage[normalem]{ulem}
\usepackage[hyphens]{url}
\usepackage[keeplastbox]{flushend}

\usepackage{array}
\usepackage[ruled,vlined]{algorithm2e}
\usepackage{comment}
\usepackage{subfig}
\usepackage{multirow}
\usepackage{booktabs}
\usepackage{soul}

\usepackage[bookmarks=true,breaklinks=true,colorlinks,linkcolor=black,citecolor=blue,urlcolor=black]{hyperref}

\def\BibTeX{{\rm B\kern-.05em{\sc i\kern-.025em b}\kern-.08em
    T\kern-.1667em\lower.7ex\hbox{E}\kern-.125emX}}

\newcommand{\headd}{{\fontfamily{qcr}\selectfont HE\_Add}}
\newcommand{\hemult}{{\fontfamily{qcr}\selectfont HE\_Mult}}
\newcommand{\herotate}{{\fontfamily{qcr}\selectfont HE\_Rotate}}

\title{Cheetah: 
Optimizing and Accelerating Homomorphic Encryption
for Private Inference}
\author{Brandon Reagen*$^{1,2}$, Wooseok Choi*$^{3}$, Yeongil Ko$^{4}$, Vincent T. Lee$^{5}$\\
        Hsien-Hsin Sean Lee$^{2}$, Gu-Yeon Wei$^{4}$, David Brooks$^{2,4}$\\
\\
*Equal contribution\\
New York University$^1$, Facebook AI Research$^2$, Seoul National University$^3$\\
Harvard University$^4$, Facebook Reality Labs$^5$}

\begin{document}
\maketitle
\pagestyle{plain}


\begin{abstract}\label{sec:abstract}
As the application of deep learning continues to grow, 
so does the amount of data used to make predictions.
While traditionally, big-data deep learning was constrained by computing performance and off-chip memory bandwidth,
a new constraint has emerged: privacy.
One solution is homomorphic encryption (HE).
Applying HE to the client-cloud model allows cloud services to 
perform inference directly on the client's encrypted data.
While HE can meet privacy constraints, 
it introduces enormous computational challenges and remains impractically slow in current systems.

This paper introduces Cheetah, a set of algorithmic and hardware optimizations for
server-side HE DNN inference to approach plaintext speeds. 
Cheetah proposes HE-parameter tuning optimization and operator scheduling optimizations, which together deliver 79$\times$ speedup over state-of-the-art. However, this still falls short of plaintext inference speeds by almost four orders of magnitude.
Cheetah further proposes an accelerator architecture, when combined with the algorithmic optimizations, to bridge the remaining performance gap.
We evaluate several DNNs and show that privacy-preserving 
HE inference for ResNet50 can be done at near plaintext performance with an accelerator dissipating 30W and 545mm$^2$ in 5nm.

\end{abstract}

\section{Introduction}\label{sec:introduction}

Deep learning lies at the heart of many modern services and applications,
and is one of the most widely used methods to process personalized data.
These models have become so successful and computationally efficient
that deep learning is now integral to everyday life.
However, as such services become ever-intricately woven into our lives, 
there is growing demand for privacy-preserving machine learning -- a
daunting task that this paper seeks to address.

Several techniques exist that offer privacy for deep learning inference that trade off the degree of security delivered versus computational efficiency.
Generally, these techniques deliver security via system implementation or mathematical guarantees.
Implementation-based methods include (i) moving computation to
edge devices, i.e., {\it local computation}~\cite{zhu2018cloud,satyanarayanan2017emergence},
and (ii) {\it trusted execution environments} (TEEs), e.g., SGX~\cite{costan2016intel,tramer2018slalom,brasser2018voiceguard}. 
Both methods achieve security by monitoring and
restricting data usage via a combination of software and hardware implementations.
In contrast, methods offering provable mathematical guarantees provide a 
theoretically-quantifiable level of privacy. 
Such solutions include
(i) {\it differential privacy} (DP)~\cite{appleLDP,ding2017collecting,erlingsson2014rappor,8416855}, 
(ii) {\it secure multi-party compute} (MPC)~\cite{rouhani2017deepsecure,liu2017oblivious,riazi2018chameleon,juvekar2018gazelle},
and 
(iii) {\it homomorphic encryption} (HE)~\cite{bos2014private,gilad2016cryptonets,hesamifard2017cryptodl,sanyal2018tapas}.
Table~\ref{tab:sec_taxonomy} summarizes the techniques and limitations associated with each with respect to wide-scale deployment.

\begin{table}[t]
\begin{center}
\begin{small}
\caption{
Generalization of privacy-preserving techniques.
}
\label{tab:sec_taxonomy}
\vspace{-0.5em}
\begin{tabular}{c c c}
\toprule
Solution    & Security       & Limitation     \\
\midrule
Local       & System         & Edge performance; leaks model   \\
TEE         & System         & Performance; side-channels     \\
DP          & Statistical    & Applications; utility-privacy tradeoff \\
MPC         & Cryptographic  & Communication bandwidth      \\
HE          & Cryptographic  & Compute        \\
\bottomrule
\end{tabular}
\end{small}
\end{center}
\vspace{-2em}
\end{table}

{
\setlength{\belowcaptionskip}{-3pt}
\begin{figure*}[t]
\begin{center}
\centerline{\includegraphics[width=\textwidth]{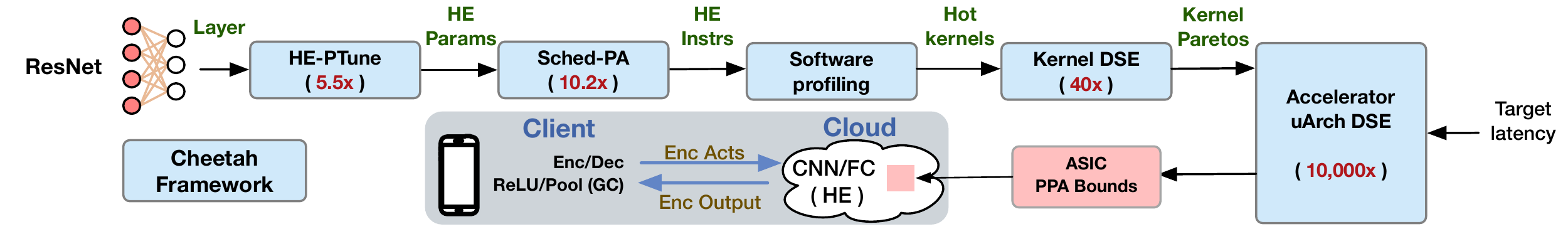}}
\vspace{-0.5em}
\caption{The Cheetah framework and system design. 
Speedup achieved for ResNet50 is reported in red.}
\label{fig:sys_cartoon}
\end{center}
\vspace{-3em}
\end{figure*}
}
Each of the above solutions have differing limitations.
Local execution offers individual users improved security, but there is risk of sensitive information leaking or being stolen through the model, plus model-privacy concerns for service providers~\cite{tramer2016stealing}.
TEEs have been shown to be vulnerable to side-channel attacks~\cite{costan2016intel}.
DP offers statistical privacy levels quantified via privacy loss $\epsilon$ but imposes an abstruse trade-off between $\epsilon$ and data utility~\cite{dwork2014algorithmic}. 
Moreover, while DP has seen success in training~\cite{chaudhuri2011sample,papernot2016semi}, its application to inference is an open question.
MPC also delivers cryptographically-strong privacy guarantees. 
However, MPC performance is limited by communication bottlenecks~\cite{rouhani2017deepsecure,liu2017oblivious,7958569}, which require consideration of network-protocol and technology levels, or reformulating the algorithm itself to alleviate.

This paper focuses on homomorphic encryption (HE) to enable privacy-preserving deep learning inference, or HE inference.
The key strength of HE is that it offers cryptographically-strong privacy guarantees, 
but these guarantees come at the cost of massive computational overheads.
These overheads are so high that
existing state-of-the-art implementations of 
HE inference~\cite{gilad2016cryptonets,hesamifard2017cryptodl,sanyal2018tapas} 
are still \emph{five to six orders of magnitude slower than plaintext inference, 
i.e., unencrypted inference speed running on a CPU}.
To put this in perspective, 
the current state-of-the-art HE inference solution (Gazelle~\cite{juvekar2018gazelle})
takes 800ms for a single MNIST inference. 
These computational overheads are so extreme that prior research has yet to
consider modern datasets and models, e.g., ImageNet and ResNet50,
as even MNIST is currently beyond the realm of feasibility.
In this paper, we propose a three-part algorithm-hardware co-design
to demonstrate that the six order-of-magnitude performance gap in the state-of-the-art can be overcome.


High-performance HE inference requires addressing three key challenges.
First, at the algorithmic level, HE has configurable parameters that 
trade performance (i.e., HE operator latency) and
``computational budget,'' 
canonically know as the \emph{noise budget} in HE literature. 
This HE noise budget limits the amount of computation (i.e., number of HE operations) that can be applied to encrypted data while still allowing correct decryption. 
Aggressive HE parameter setting improves performance by reducing the cost of each operation
(e.g., using smaller data types),
but if set too aggressively, the noise budget can be exceeded and
cause the computation (i.e., decryption) to fail. 
The second challenge is how computations are {\it scheduled} and mapping to HE primitives.
HE only supports a limited set of operators (e.g., add and multiply) that applications
must be expressed as, and each operator increases noise differently.
Therefore, noise-aware operator schedules can significantly improve performance 
by reducing accumulated noise,
enabling more aggressive HE parameters to be used.
The final challenge is the sheer number of computations HE inference entails.
As we show, this challenge requires hardware acceleration and leveraging
the extreme degrees of parallelism in both DNNs and HE operators to maximize performance.



\noindent \textbf{Key Contributions:} 
To address these challenges, this paper presents \emph{Cheetah}, a framework (Figure~\ref{fig:sys_cartoon}) to enable practical HE-based privacy-preserving machine learning inference by combining algorithm optimization and custom hardware acceleration. 
We assume Gazelle~\cite{juvekar2018gazelle}, the state-of-the-art, as our baseline.
Our contributions are as follows:

First, \textbf{we propose \emph{HE-PTune} (Section~\ref{sec:models}), which is an analytical model that tunes HE parameters} at the algorithm level.
HE-PTune automatically identifies the highest-performance HE parameter settings that satisfy noise-budget constraints by tuning HE parameters based on the needs of each layer in a deep neural network model.
HE-PTune's parameter tuning yields up to 11.7$\times$ for VGG16 and 5.5$\times$ for ResNet50 performance benefit over state-of-the-art.

Second, \textbf{we propose a new schedule for dot product operations called \emph{Sched-PA} to minimize consumption of the noise budget and improve performance in HE}.
Sched-PA is a partial-aligned dot product schedule, which exploits the key insight that the order of HE operations significantly impacts performance and noise budget.
This allows Sched-PA to achieve a maximum additional speedup of 10.2$\times$ (5.2$\times$ harmonic mean) and a combined speed using HE-PTune of 79.6$\times$ (13.5$\times$ harmonic mean) over state-of-the-art.


Third, \textbf{we propose a custom hardware accelerator architecture that combines these algorithmic optimizations to accelerate server-side HE inference to approach plaintext speeds} given the abundance of parallelism and opportunities for specialization.
To do this, we first conduct hot kernel profiling of an HE CPU software implementation~\cite{sealcrypto} to derive the speedups necessary for plaintext latency using parameters identified by HE-PTune and Sched-PA. 
We also identify the amount of application inter-kernel and intra-kernel parallelism available in hot kernels.
We then use these profiling results to implement a custom accelerator architecture for HE inference and conduct design space exploration for each HE kernel to measure speedups afforded by exploiting exposed parallelism.

\textbf{Combining algorithmic optimizations with custom hardware acceleration, Cheetah approaches speeds comparable to plaintext inference.}
Cheetah is \emph{five to six orders of magnitude faster} than Gazelle.
For ResNet50, we find accelerator hardware requirements on the order of 545mm$^2$ and 30W in a 5nm technology node (Section~\ref{sec:limits}). 
More importantly, we find that the accelerator area and power resources required to support HE inference at these speeds is within practical (albeit still high) resource requirements which is still on the order of a large datacenter-class GPU or similar coprocessor.
\section{Overview and Assumptions}\label{sec:overview}

{
\begin{figure*}[t]
\begin{center}
\centerline{\includegraphics[width=\textwidth]{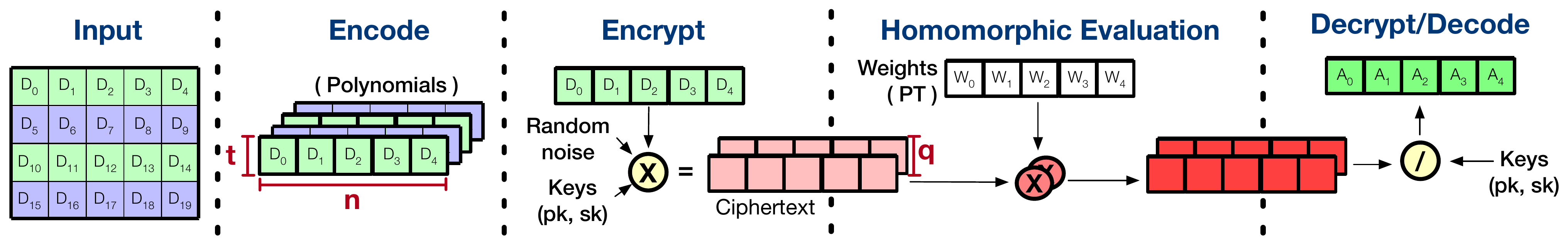}}
\vspace{-1em}
\caption{Overview of how data is processed using BFV homomorphic encryption.}
\label{fig:bfv_cartoon}
\end{center}
\vspace{-3em}
\end{figure*}
}

\subsection{System Setup}\label{sec:sys_assumption}
A typical deep learning system setup is shown in the gray box of Figure~\ref{fig:sys_cartoon}.
A client generates data and sends it to the cloud.
The cloud performs inference and the result is returned to the client.
The most direct way to apply HE is for the client
to encrypt the data, the cloud processes the entire inference using HE,
and the encrypted result is returned to the client.
Unfortunately, this approach has two drawbacks: 
(1) HE cannot readily process nonlinear functions (without incurring prohibitively large penalties)
and (2) many computations in DNNs requires a relatively large HE noise budget, 
which necessitates larger encryption parameters, resulting in poor 
performance. This effect is exacerbated by deeper networks.

A classic solution is to combine
multiple cryptographic solutions, as done before in~\cite{malkhi2004fairplay,orlandi2007oblivious,sadeghi2009efficient,7958569,liu2017oblivious},
and partition inferences across them.
The typical approach is to execute linear operators (FC/CNNs) on the cloud
using either homomorphic encryption (Gazelle~\cite{juvekar2018gazelle}) or secret-sharing (MiniONN~\cite{liu2017oblivious}),
and non-linear functions on the client with Yao's garbled circuit (GC).
This partition works quite well as GCs incur small computational overhead~\cite{bellare2013efficient,zahur2015two} while the cloud can leverage powerful
servers to handle the large processing load of HE/secret sharing.
In this paper, we take Gazelle as our baseline as it is the fastest known implementation
of private inference and uses HE, which is the focus of Cheetah.


In Gazelle, the client encrypts data to be processed and sends it to the cloud.
The cloud applies a single linear layer (e.g., convolution) to the input using HE.
ReLU and pooling functions are computed on the client using a GC.
The GC is configured by the cloud and sent to the client along with the encrypted linear layer outputs.
The client then decrypts the outputs and processes them using the GC.
Note that allowing the client to observe the original outputs 
after decryption can leak the cloud's private model weights
(knowing the inputs and outputs of a linear function would
make it trivial to steal the cloud's model).
To prevent this, the cloud obscures the actual activation values (both input and output)
by adding random numbers to each, i.e.,
the client receives encrypted activation input also obfuscated with random numbers.
After decryption, the client runs the GC, 
which includes a subtraction circuit to remove the added random numbers securely
(recovering original values), 
an non-linear functions (ReLU or pooling), 
and finally an addition to obscure the plaintext output value and protect model weights.
Once GC evaluation completes, the masked output is re-encrypted by the client and
sent to the cloud.
On the cloud, the random numbers added to the activation are removed via HE subtraction
and the following linear layer is computed (using HE).
The HE-MPC cycle repeats for each layer of the deep network.

Note that in homomorphic encryption, decryption resets the HE noise budget.
Therefore, systems like Gazelle address both issues associated with nonlinear computation and limitations of HE noise budget.
However, the computational overheads of HE---the
focus of this paper---remain prohibitive.
Cheetah addresses the HE compute bottleneck, which is an architecture/hardware
problem, but the proposed optimizations for HE are more generally applicable to other solutions beyond Gazelle.
Solving the communication/network bottleneck is beyond the scope of this paper.
We expect contributions on the algorithmic 
(e.g., different MPC protocols~\cite{mohassel2018aby,orlandi2007oblivious,fenner2020privacypreserving}) 
and technology (e.g., 5G) front to help.
Therefore, Cheetah assumes the same communication overheads as Gazelle.
Whenever discussing HE performance results, 
it is always with respect to the server-side HE inference computation.


\subsection{Threat Model}
The threat model assumed by Cheetah is the same as in Gazelle~\cite{juvekar2018gazelle},
and similar to other two-party compute (2PC) solutions including
DeepSecure~\cite{rouhani2017deepsecure}, MiniONN~\cite{liu2017oblivious}, and SecureML~\cite{7958569}.
The model assumes the client/user and cloud are honest but curious, i.e.,
each agent follows the protocol precisely but may try to infer information. 
Under this assumption, Cheetah 
preserving the privacy of both the clients' data and cloud's model weights.
For more details, see~\cite{juvekar2018gazelle}.

Note that the protocol {\it does} leak some information about the model.
Because ReLU and pooling layers are performed by the client,
the client can learn the number and shape of each layer. 
The model weights values, however, are not leaked.
It is possible to obscure this information 
(e.g., pad tensor dimensions and add null layers),
but they are not considered here and left as future work.
Cheetah focuses on improving users privacy while 
protecting the cloud's models (considered IP today~\cite{zhang2018protecting})
from model-stealing attacks~\cite{tramer2016stealing}.

\section{Background}\label{sec:background}

This section provides a brief introduction to HE
and the BFV~\cite{fan2012somewhat}. 
For a complete description see \cite{brakerski2012fully, fan2012somewhat}.

\subsection{Homomorphic Encryption: The Basics}

HE is a privacy-preserving encryption technique that enables computation over encrypted data,
which was first shown to be possible by Gentry~\cite{Gentry:2009:FHE:1536414.1536440}.
Since its discovery, many algorithmic improvements have been made to improve performance~\cite{Gentry:2010:CAF:1666420.1666444,brakerski2014leveled,gentry2012fully,gentry2013homomorphic,bos2013improved,brakerski2012fully,fan2012somewhat}.
Modern HE schemes such as BFV allow adds and multiplies between encrypted data 
and derive security from the hardness of the Ring Learning With Error (RLWE) problem~\cite{lyubashevsky2010ideal}.
In BFV, 
noise is added during plaintext encryption and accumulates over successive ciphertext computations.
If the aggregate noise exceeds a noise budget threshold, decryption fails.
This noise budget is a function of the HE parameters and defines how many computations can occur before decryption fails.
HE schemes of this type are called {\it Leveled} HE (LHE).
In contrast, fully homomorphic encryption (FHE) schemes enable an arbitrary number of computations.
FHE schemes can be built from LHE schemes via \textit{bootstrapping}~\cite{Gentry:2009:FHE:1536414.1536440,fan2012somewhat}.
Bootstrapping reduces the noise in the ciphertext but is expensive to implement, so most applications focus on LHE.

\begin{table}[t]
\begin{center}
\begin{small}
\caption{BFV parameters.}
\label{tab:bfv_params}
\vspace{-0.5em}
\begin{tabular}{cc}
\toprule
Parameter &  Description \\
\midrule
$n$   & Polynomial degree (vector length) \\
$t$   & Plaintext (pt) modulus \\
$q$   & Ciphertext (ct) modulus \\
$W_{dcmp}$ & Weight (pt) decomposition base \\
$A_{dcmp}$ & Activation (ct) decomposition base \\
$\sigma^2$   & Variance of noise added for encryption (fixed) \\
\bottomrule
\end{tabular}
\end{small}
\end{center}
\vspace{-2.5em}
\end{table}

\subsection{BFV: Relatively Efficient HE}

BFV~\cite{fan2012somewhat} is a relatively efficient LHE scheme; 
Figure~\ref{fig:bfv_cartoon} shows an overview of the process.
In BFV, data is \textit{encoded} as a plaintext polynomial 
that is then \textit{encrypted} as a pair of ciphertext polynomials.
Ciphertexts are then input to addition and multiplication during \textit{evaluation}.
The resulting ciphertexts from evaluation are \textit{decrypted} to plaintext and finally \textit{decoded}
to individual scalars.
Polynomials are implemented as integer vectors, where the vector length (polynomial degree) 
and bit-width (coefficient size) are set by HE parameters.
BFV parameters (listed in Table~\ref{tab:bfv_params}) must be carefully tuned as they affect
computational efficiency and security.

\textit{Core BFV Parameters ($n$, $t$, $q$):}
Plaintext polynomials are elements of the ring: $R_t = \mathbb{Z}_t[x]/(x^n+1)$, 
where the degree of the polynomial is less than $n$ (a power of 2).
Polynomial coefficients are integers in $\mathbb{Z}_t$
(integers in the range $(-\frac{t}{2}, \frac{t}{2}]$).
$t$ is called the \textit{plaintext modulus} as all HE operations
are taken modulo $t$ in the plaintext space.
Setting $t$ requires profiling the application 
to ensure enough bits are used for correctness and no more,
as over provisioning causes unnecessary slowdown.

Similarly, two polynomials of a ciphertext are in $R_q = \mathbb{Z}_q[x]/(x^n+1)$,
where $q$ is the \textit{ciphertext modulus}.
The ratio between $q$ and $t$ determines the noise budget, 
which sets the number of HE operators that can 
be computed per ciphertext before decryption fails.
The ratio between $n$ and $q$ 
for a given variance ($\sigma^2$) of Gaussian noise added for encryption sets the security strength of the HE scheme (see~\cite{fan2012somewhat} for details).

\textit{Encoding (Packing) Data to Polynomial:}
When proper HE parameters are used 
(i.e., $t$ is prime and $t \equiv 1 \mod 2n$),
a property of the ring $R_t$ enables a form of algorithmic parallelism.
Here, each plaintext polynomial in $R_t$ and, hence, the ciphertext, can be \emph{packed} with $n$ data.
This means that each HE addition or multiplication can actually
perform an $n$-way parallel element-wise computation.
With packing, each scalar data is tied to a \textit{slot}, 
and slots can be thought of as individual elements in the integer array.
Packing significantly improves HE performance;
$n$ is typically on the order of thousands 
and the benefits of packing are proportional~\cite{smart2014fully}.

\textit{Polynomial Representations:}
Polynomials are represented in two spaces---coefficient and evaluation.
The coefficient representation is how polynomials are typically represented,
e.g., $\sum_{i=0}^{n-1} \alpha_ix^{i}$.
The evaluation space is analogous to
the frequency domain of time-domain signals.
Similar to FFT, efficient conversion between the two is done via the
Number Theoretic Transform (NTT)~\cite{brakerski2014leveled,smart2014fully}.
Cheetah keeps polynomials in the evaluation space and
converts to coefficient space only as needed for operations like decomposition (see below).
Using the evaluation space as a default representation reduces the number
of NTTs needed for homomorphic CNN/FC.
Note that applying NTT to ciphertexts does not affect noise.

\begin{table}[t]
\begin{center}
\begin{small}
\caption{Impact on Noise of basic BFV operations.}
\label{tab:noise_impact}
\vspace{-0.5em}
\begin{tabular}{cc}
\toprule
 &  Noise Bound after Each Operation \\
\midrule
Noise $(v_0)$ in fresh $\textrm{ct}_0$  & $2nB^2\,\, (B=6\sigma)$ \\
\headd$(\textrm{ct}_0, \textrm{ct}_1)$ & $v_0 + v_1$ (additive) \\
\hemult$(\textrm{pt}, \textrm{ct}_0)$ & $ nl_{pt}W_{dcmp}v_0/2$ (multiplicative) \\
\herotate$(\textrm{ct}_0)$ & $ v_0 + l_{ct}A_{dcmp}Bn/2$ (additive) \\
\bottomrule
\end{tabular}
\end{small}
\end{center}
\vspace{-1.5em}
\end{table}

\subsubsection{Operations of BFV}\label{sec:bg_hebfv}

BFV consists of three operators: \headd, \hemult, and \herotate.
Recall that the \headd~and \hemult~operate on vectors of packed data,
so they are effectively SIMD-add and SIMD-multiply operations.
Note that the underlying implementations of \headd~and \hemult~consist 
of many modular arithmetic calculations, different from a single cycle integer add or multiply computation.
Table~\ref{tab:noise_impact} shows the amount of noise introduced by each operator which depend on BFV parameters.
$B$ is the bound of the noise added during encryption while $v_i$ represents the noise in ciphertext ct$_i$.
The remaining parameters ($l_{pt}$, $l_{ct}$, $W_{dcmp}$, and $A_{dcmp}$) are for decomposition,
defined in Section~\ref{label:decomposition}.



\headd: Two ciphertexts can be added homomorphically by 
summing each ciphertext coefficient followed by a modulo operation.
I.e., a resulting coefficient outside the range $\mathbb{Z}_q$ is reduced to be in $\mathbb{Z}_q$.
Reduction is implemented as a comparison and subtraction to keep the performance overhead low.
Each \headd~operation increases noise additively.

\hemult: BFV supports both ct-ct and pt-ct multiplication.
Cheetah uses pt-ct multiplication to multiply plaintext weights by encrypted activations. 
Pt-ct multiplication is achieved by multiplying evaluation space ciphertext polynomials by 
the evaluation space plaintext polynomial containing weights on a per-element basis.
Performance is limited by the modular reduction required for each polynomial coefficient of output.
Cheetah uses Barret reduction (see Section~\ref{perf_model}).
\hemult~operations increases noise multiplicatively.

\herotate: BFV supports slot rotation within a packed polynomial to 
enable computation between data in different slots.
Since \headd~and \hemult~are element-wise operations, 
computations like dot products require \herotate~to align partial products and implement 
the reduction (see Section~\ref{sec:partial_align}).
\herotate~is computationally expensive with many steps, and increases noise additively.
We refer the reader to~\cite{brakerski2014leveled,wu2012using} for details.


\begin{table}[t]
\begin{center}
\begin{small}
\caption{
HE-PTune performance models.
}
\label{tab:perf_model}
\begin{tabular}{ccc}
\toprule
CNN & \hemult & \herotate  \\
\midrule
$n \geq w^2$ & $l_{pt}c_ic_of_w^2/c_n$ & $c_ic_of_w^2/c_n$  \\
$n < w^2$ & $l_{pt}(2c_n-1)c_ic_of_w^2$ & $(2c_n-1)c_ic_o(f_w^2-1)$  \\
\midrule
FC & \hemult & \herotate \\
\midrule
$n \geq n_i, n \geq n_o$ & $l_{pt}n_in_o/n$ & $n_in_o/n -1 + \log (n/n_o)$  \\
$n \geq n_i, n < n_o$ & $l_{pt}n_in_o/n$ & $(n_i-1)n_o/n$  \\
$n < n_i, n \geq n_o$ & $l_{pt}n_in_o/n$ & $(n_o + \log (n/n_o))n_i/n$  \\
$n < n_i, n < n_o$ & $l_{pt}n_in_o/n$ & $(n-1)n_in_o/n^2$  \\
\bottomrule
\end{tabular}
\end{small}
\end{center}
\vspace{-3em}
\end{table}

\subsubsection{Polynomial Decomposition}\label{label:decomposition}

Decomposition is used to segment polynomials into multiple components with smaller-valued coefficients.
The key idea is that HE operations over smaller coefficient polynomials reduces noise growth.
To enable this, Cheetah has two parameters for polynomial decomposition: $W_{dcmp}$ and $A_{dcmp}$
(Table~\ref{tab:bfv_params}), which defines the base that polynomials are decomposed to.
Decreasing decomposition base increases the number of decomposed polynomials
which decreases operator noise growth but increases the total amount of compute.
Once decomposed operators complete, resulting segments are summed to get the final result.

\herotate\, requires ciphertext decomposition, otherwise a single operation can exceed the noise budget.
The decomposition base $A_{dcmp}$ is used to factor ciphertext polynomials into multiple,
smaller-magnitude polynomials when \herotate~is applied.
We denote $l_{ct} \approx \log_{A_{dcmp}}(q)$ as the number of polynomials with base $A_{dcmp}$ resulting from the decomposition.
Since \herotate~noise increase is additive, with decomposition noise increase by an additive factor proportional to $A_{dcmp}$ and the increase in number of polynomial operations $l_{ct}$.


\hemult\, also benefits from decomposition to reduce noise.
For neural networks, we use \hemult~with decomposition to compute the partial products since weights are presented in plaintext.
Using a decomposition base $W_{dcmp}$, the plaintext polynomial can be decomposed into $l_{pt} \approx \log_{W_{dcmp}} (t)$ polynomials.
The resulting \hemult~with decomposition requires $l_{pt}$ polynomial multiplications to implement but reduces noise growth by a factor of around $t/(l_{pt}W_{dcmp})$.

\section{HE-PTune: Models \& Parameter Tuning}\label{sec:models}

HE parameter selection is a major source of complexity (i.e., $n, t, q, W_{dcmp}$, $A_{dcmp}$),
where proper selection strikes a balance between noise budget and performance.
A greater noise budget enables more computations per ciphertext but slower HE operators.
Existing solutions rely on over-provisioning noise budgets,
resulting in suboptimal performance.
This section proposes HE-PTune: analytical performance and noise models for HE DNN operators to maximize performance via fine-grained parameter tuning.
Tuning parameters with HE-PTune delivers up to a 11.7$\times$ speed up over the state-of-the-art.

\begin{table}[t]
\begin{center}
\begin{small}
\caption{
Noise models for CNN and FC layer. 
}
\label{tab:noise_model}
\vspace{-0.5em}
\begin{tabular}{cc}
\toprule
CNN & Output Noise  \\
\midrule
$n \geq w^2$ &  $f_w^2c_i\eta_Mv_0 + \eta_Ac_i(f_w^2-1+(c_n-1)/c_n)$  \\
$n < w^2$ &  $(2f_w-1)f_wc_i\eta_Mv_0 + \eta_Ac_i(2f_w+1)(f_w-1)$  \\
\midrule
FC & Output Noise  \\
\midrule
$n \geq n_i$ & $n_i\eta_Mv_0 + \eta_A(n_i-1)$   \\
$n < n_i$ & $n_i\eta_Mv_0 + \eta_An_i(n-1)/n$   \\
\bottomrule
\end{tabular}
\end{small}
\end{center}
\vspace{-3em}
\end{table}

\subsection{Performance Modeling}\label{perf_model}

HE-PTune's performance model analytically derives the total number of
underlying integer multiplications per layer.
(Recall that the HE operator \hemult~consists of many integer multiplications.)
Most HE operators resolve to multiplication operations
and ones that do not have run-times either strongly correlated or dominated by those that do.
Performance models for CNN and FC layers are built by
first capturing all HE and NTT operations.
Then all operations are reduced to the total number of underlying integer multiplication operations.

\begin{figure*}
\centering
\subfloat[Layer5 (FC)]{\includegraphics[width=0.33\textwidth]{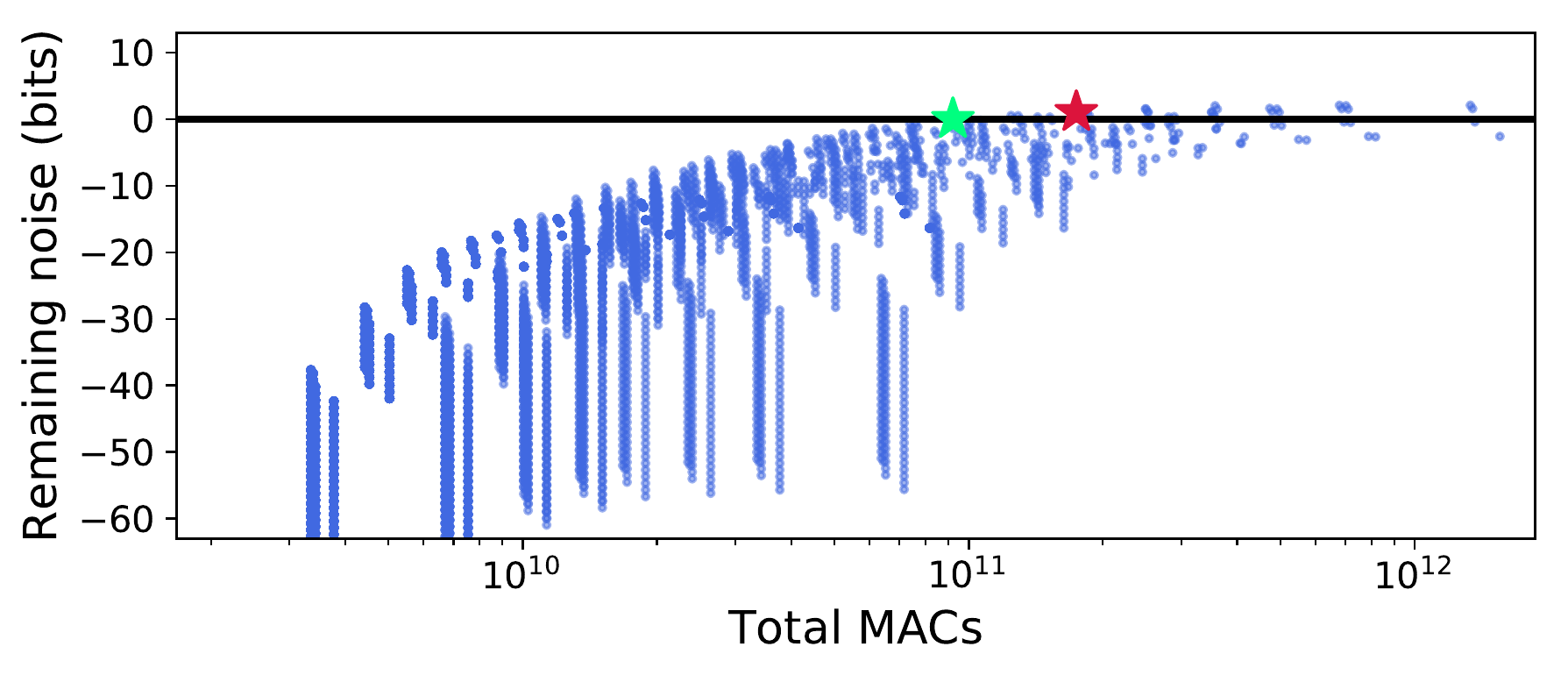}}
\subfloat[Layer0 (CNN)]{\includegraphics[width=0.33\textwidth]{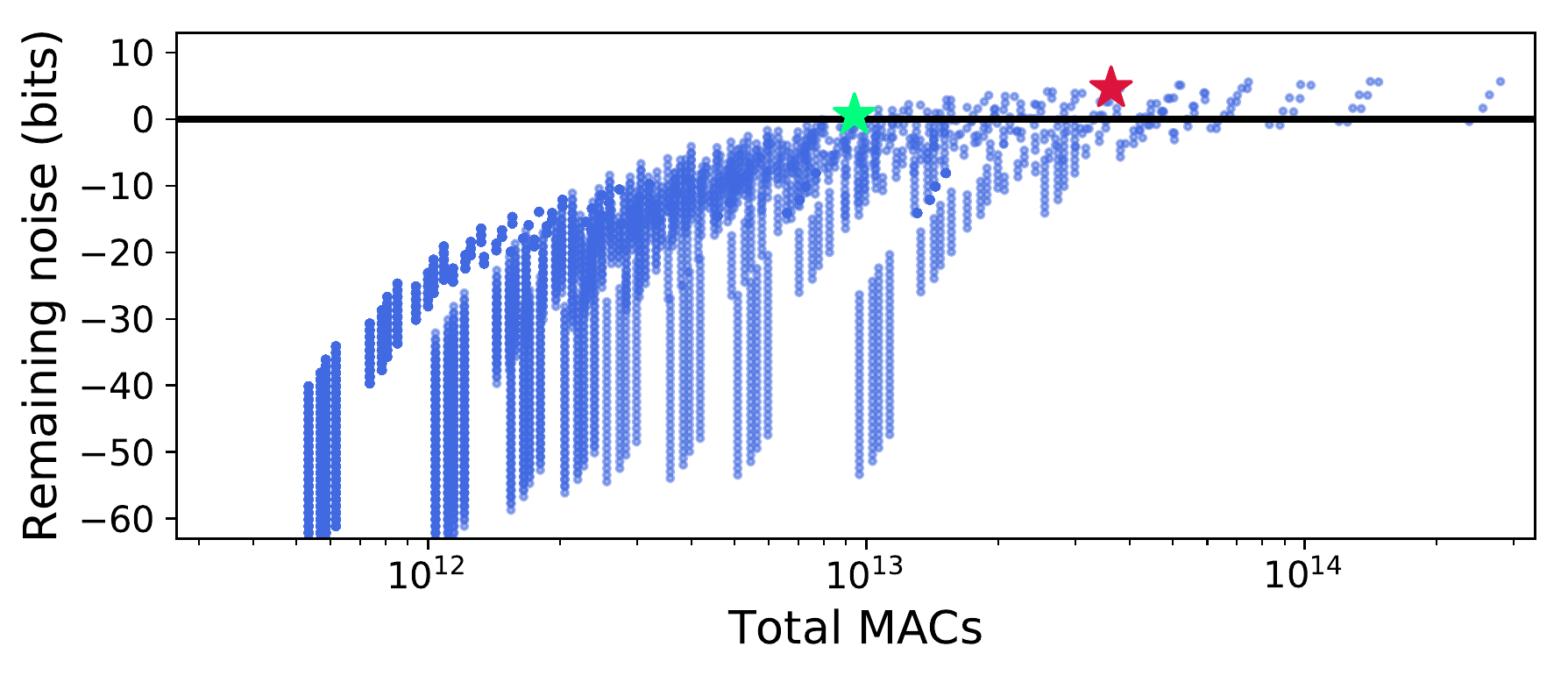}}
\subfloat[Alexnet Speedup]{\includegraphics[width=0.33\textwidth]{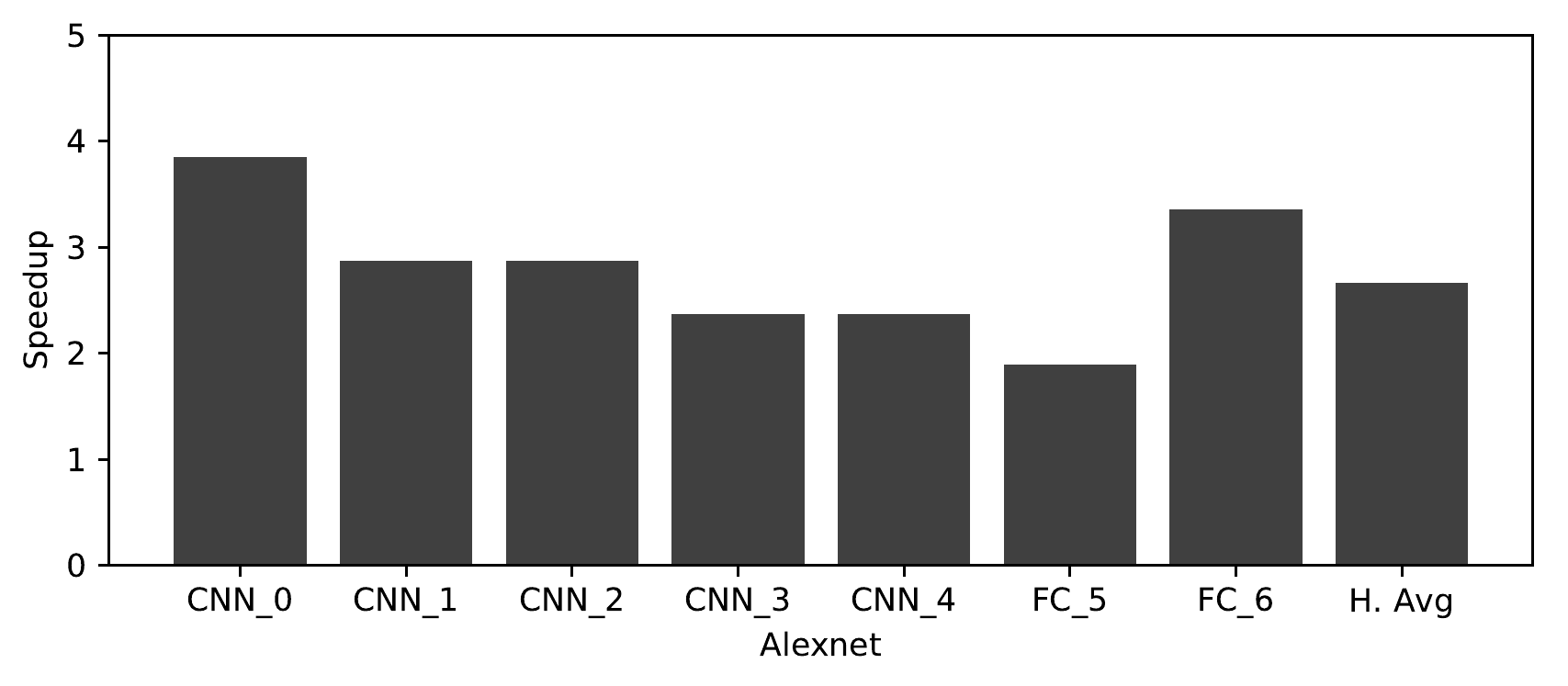}}
\vspace{-.5em}
\caption{
Comparison of HE-PTune and Gazelle using AlexNet.
Blue dots are HE configurations modeled by HE-PTune. 
The red star is Gazelle's configuration and the green star is the optimal found by HE-PTune.
Layer5 and Layer0 show the best and worst configuration 
for Gazelle with respect to utilized noise budget.
HE-PTune's speedup for all layers on the right.}
\label{fig:opt_dse_across}
\vspace{-1.5em}
\end{figure*}

\subsubsection{Modeling CNNs}

CNN layers are parameterized as $(w, f_w, c_i, c_o)$,
where $w^2$ and $f_w^2$ represent the size of input image and weight filter,
and $c_i$ and $c_o$ denote the number of input and output channels.
Encryption parameters follow the notation defined in Table~\ref{tab:bfv_params}.
Effective modeling of \hemult~and \herotate~counts require consideration of two cases: 
1) the ciphertext slot count is greater than an input image (i.e., $n \geq w^2$),
2) the ciphertext slot count is less than an input image (i.e., $n < w^2$).
We use $c_n$ to model the relationship between ciphertext slots and input image size.
$c_n$ is defined as
the number of input image channels per ciphertext (i.e., $n/w^2$) in the first case
and the number of ciphertexts per input image channel ($w^2/n$) in the second.
Table~\ref{tab:perf_model} shows how each case
counts the number of HE operations per CNN layer.

\herotate~operations require both polynomial multiplication and NTTs. 
Precisely, assuming a ciphertext decomposition base $A_{dcmp}$, 
$2l_{ct}$ multiplications and $l_{ct}+1$ NTT ($l_{ct} \approx \log_{A_{dcmp}}q$) are required per \herotate.
Each $n$-point NTT entails $n\log n/2$ butterflies. 
Cheetah uses Harvey's butterfly (3 integer-multiplications per butterfly).
\hemult~does not require NTTs as  
in Cheetah the default polynomial representation is the evaluation space.
Each \hemult~requires two element-wise modular multiplications between the two polynomials, resulting in $2n$ modular multiplications per \hemult.
Cheetah uses Barrett reduction~\cite{katz1996handbook},
which uses five integer-multiplications per reduction.

\subsubsection{Modeling FCs}
A similar process is repeated to model FC layers.
The required number of integer multiplications per
\hemult~and \herotate~operations is the same in both CNN and FC,
the only difference is the number of \hemult~and \herotate~counts.
Here, an FC layer is parameterized as $(n_i, n_o)$, 
where $n_i$ and $n_o$ represent the number of input and output activations.
The required number of \hemult~and \herotate~for all possible cases
are summarized in Table~\ref{tab:perf_model}.

\subsection{Noise Modeling}

Once CNN/FC layers are implemented as HE operations (see Section~\ref{sec:partial_align}),
noise growth can be modeled using the equations in Table~\ref{tab:noise_impact}.
We developed a model for layer noise as a function of both
HE ($n, t, q, W_{dcmp}, A_{dcmp}$) and DNN ($f_w, w, c_i, c_o$ for CNN and $n_i, n_o$ for FC) parameters.
Note that directly applying the equations in Table~\ref{tab:noise_impact}
estimates noise in the \emph{worst} case. 
The worst case is very rare (see below) and causes unnecessarily slow HE parameters to be selected.
To overcome this, we develop practical noise estimations for HE operators and
provide a theoretical analysis of the decryption failure rate.
We also note that all prior work on high-performance
HE~\cite{gilad2016cryptonets,brutzkus2019low,juvekar2018gazelle} 
sets HE parameters using heuristics, 
providing high-likelihoods for success but not guaranteeing it.

Cheetah builds a theoretically-motivated, empirically-derived noise model
that minimizes computational overheads for a targeted probability of success.
We observe that added encryption noise
is sampled from an independent bounded discrete Gaussian (IBDG) distribution with variance $\sigma^2$,
and if $X_i$'s are IBDG with variance $\sigma_i^2$, 
then $\sum_i \alpha_i X_i$ is also IBDG
with variance $\sum_i \alpha_i^2 \sigma_i^2$.
As the noise grows multiplicatively in \hemult~and additively in \headd\, and \herotate, 
we can compute the variance of the output noise 
after each layer under the independence assumption,
which was validated in \cite{ducas2015fhew}.
Then, since the output noise ($Y$) is IBDG with standard deviation ($\sigma_Y$),
the probability of decryption failing is bound by 
$\textrm{Pr}(|Y|\geq q/(2t)) \leq 2\exp{(-q^2/(4t^2\sigma_Y^2))}$.
We use these equations to derive an output noise threshold for a probability of correct decryption.
Therefore, instead of using worst-case bounds and guaranteeing correct decryption,
our noise model uses the scaled expressions given in Table~\ref{tab:noise_impact}.
Cheetah uses a scaling factor $c$ such that the decryption failure rate is provably less than $10^{-10}$, which is negligible as it is much lower than the DNN's misclassification rate.



The noise models are given in Table~\ref{tab:noise_model}. 
Here, $v_0$ is the initial noise for the input ciphertext,
$\eta_{M}$ is the noise due to \hemult,
and $\eta_{A}$ is the growth factor from \herotate.
By dividing $\frac{q}{2t}$ by the output noise (and taking the log),
the remaining noise budget in bits is given.
When the budget is negative, decryption fails;
when positive, it fails with probability $\leq$10$^{-10}$.

\subsection{HE Parameter Space Exploration}

Using a single set of HE parameters for all DNN layers results in poor performance,
as HE parameters are provisioned for the worse case layer noise.
Using HE-PTune's models for noise and performance,
parameters can be readily tuned on a per-layer basis.
HE-PTune takes layer hyperparameters as input
and outputs optimal HE parameters found via a design space exploration.
Because the model is analytical, a vast
parameter space can be explored in a matter of minutes.

An example of HE parameter space exploration are given in Figure~\ref{fig:opt_dse_across}
for AlexNet on ImageNet.
Each blue dot is unique set of HE parameters modeled with HE-PTune
to estimate computation and remaining noise budget.
Red stars indicate parameters used by Gazelle
and green stars show the optimal point found using HE-PTune.
Gazelle uses the same sets of HE parameters for all layers.
Of all layers in the model, Layer 5 has the smallest remaining noise budget,
and it follows that the speedup between Gazelle and Cheetah is the lowest for this layer
(see bars in Figure~\ref{fig:opt_dse_across}).
Using HE-PTune, empirical results show using a 
single set of parameters is inefficient and unnecessary.
The highest Cheetah speedup is in Layer 0, 
where Gazelle has an excess noise budget of 4.6 bits 
whereas HE-PTune finds a configuration leaving only 1 bit of noise budget.
Improvements come from tailoring parameters to the requirements of each layer.

\begin{figure*}[t]
\begin{center}
\includegraphics[width=0.97\textwidth]{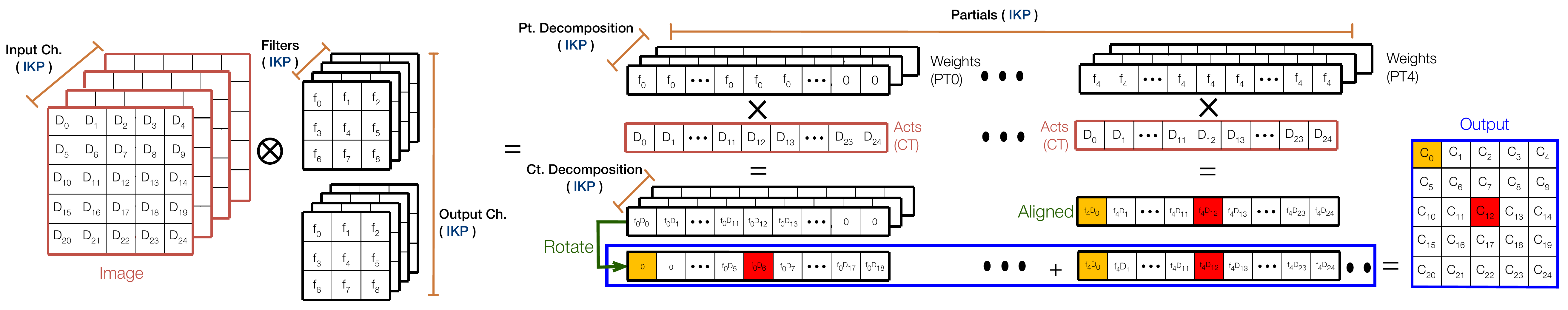}
\vspace{-0.5em}
\caption{
How Cheetah implements CNNs using Sched-PA.
Sources of inter-kernel parallelism (IKP) are labeled.
}
\label{fig:cnn_example}
\end{center}
\vspace{-2em}
\end{figure*}

\begin{figure}[t]
\begin{center}
\includegraphics[scale=0.4]{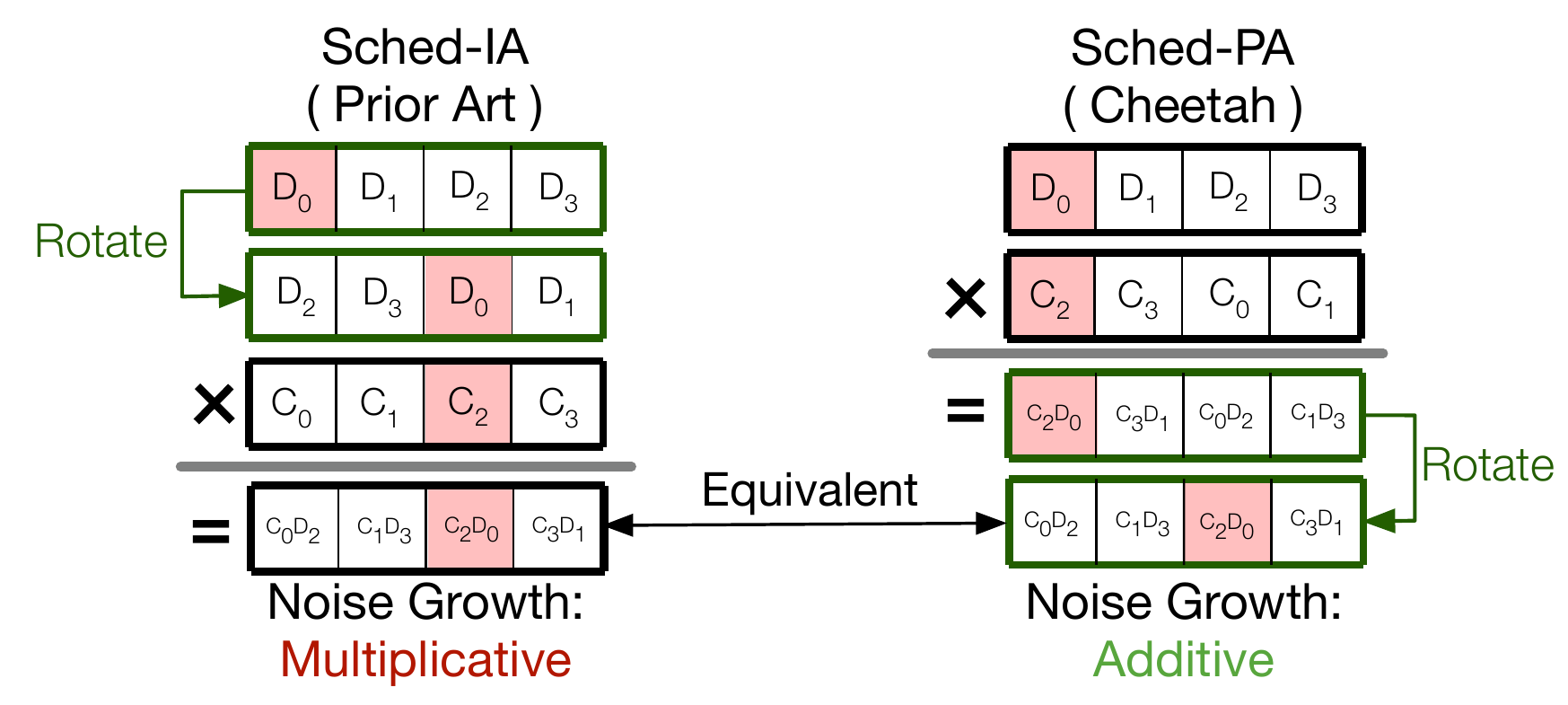}
\vspace{-0.5em}
\caption{
Sched-IA (input-aligned) versus Sched-PA (partial-aligned) dot product schedules. Cheetah uses
Sched-PA to improve performance of CNN and FC layers.
}
\label{fig:op_order}
\end{center}
\vspace{-2em}
\end{figure}

HE-PTune also eases the finding
functional HE parameter settings in the first place.
Recall that any point where the noise budget is exceeded fails to decrypt.
Of all the points evaluated in the design space search, 
over 99\% have a negative remaining noise budget and will not work.
Finding HE parameters is difficult,
and further motivating HE-PTune.

We validate the parameter sets from HE-PTune using different CNN and FC layers
used in popular DNNs, including: LeNet-300-100 and LeNet5 for MNIST~\cite{lecun1998gradient},
and AlexNet~\cite{krizhevsky2012imagenet}, VGG16~\cite{simonyan2014very}, and ResNet50~\cite{he2016deep} for ImageNet~\cite{russakovsky2015imagenet}.
Each layer is tested using a variety of HE parameters, with no consideration of noise budget to
explore the parameter space.
Execution times are collected by implementing each CNN/FC
layer in the SEAL HE library~\cite{sealcrypto} and measuring its performance on a Xeon server.
The remaining noise budget is collected after each run using
SEAL's internal measuring capability and API.
Overall, we find that due to the underlying randomness of the noise, 
the noise model shows slightly larger error than the performance model.
However, this is acceptable as the worst-case errors are within 1\,bit in the low-remaining noise budget region of the space.

\section{Partial-aligned scheduling}\label{sec:operator_optimization}

This section introduces a new dot product schedule - Sched-PA - to improve HE performance on FC and CNN layers.
Recall that each HE primitive has different run-time and additive noise trade-offs (Section~\ref{sec:background})
and the overheads of different primitive schedules are not associative so order of operations matters.
Operation orderings with less noise are beneficial as it 
enables higher-performance via more computationally efficient HE parameters.
As a result, Cheetah's Sched-PA dot product optimization provides a performance benefit of up to 10.2$\times$ compared to prior work.

\subsection{Sched-PA: Partial-Aligned Dot Products}\label{sec:partial_align}

The key challenge for implementing HE dot products is optimizing how data is
packed into polynomial slots and the relative order of operations.
Computing a dot product in HE requires all three primitives: \hemult, \headd, and \herotate.
Partial products are computed using an \hemult~operation between
a ciphertext (encrypted activation) and a plaintext (model weights).
Each partial is accumulated with a series of \headd~operations
to reduce the final output.
The challenge is that HE operations only support computation between \emph{aligned}
polynomial slots.
This means that when polynomial $\mathrm{A}$ and $\mathrm{B}$ are multiplied (resulting in $\mathrm{C}$),
$\mathrm{C[i]=A[i]\times B[i]},$ $\forall i \in [0, n)$.
To properly reduce each of the partials of a dot product, 
the slots in $\mathrm{C}$ must be aligned to use the correct values.

Prior work aligns the inputs before performing multiplication,
referred to here as an \emph{input-aligned} schedule (Sched-IA)~\cite{halevi2014algorithms, juvekar2018gazelle}.
In Sched-IA, the input ciphertext is first aligned, or rotated,
to the correct output slot, and plaintext weights are packed appropriately.
The post-rotation ciphertext and plaintext are then
multiplied, resulting in a dot product partial (ciphertext).
Resulting partial ciphertexts can be readily accumulated to compute the final value.

Cheetah proposes a new dot product implementation called \emph{Sched-PA} (see Figure~\ref{fig:cnn_example}, \ref{fig:op_order}).
Our key insight is that \hemult~increases noise
by a multiplicative factor $\eta_M$ ($\leq nl_{pt}W_{dcmp}/2$)
whereas \herotate~is additive $\eta_A$.
In Sched-PA, 
the initial input ciphertext is always kept in its original order.
Weights are again packed into a plaintext polynomial and aligned with ciphertext
slots to compute the correct partial product via \hemult.
Finally, resulting partial product ciphertexts are
aligned such that the partial slot matches the correct output slot.
Figure~\ref{fig:op_order} also shows Sched-PA compared to the other approach.

The benefit stems from noise accumulation in chained HE operations.
Recall that $v_0$ and $\eta_A$ represent the initial input ciphertext noise and additive noise from \herotate, respectively.
Thus, a dot product using the partial aligned schedule
experiences a noise growth of $\eta_Mv_0+\eta_A$.
In contrast, the Sched-IA dot product first rotates \textit{then} multiplies,
resulting in noise growth of $\eta_M(v_0+\eta_A)$, 
significantly larger than Sched-PA.
Saving noise enables HE-PTune to identify higher performance HE parameter settings,
ultimately resulting in performance benefit.

\subsection{Implementing Low-Noise Convolution}\label{sec:ln_cnn}

Figure~\ref{fig:cnn_example} shows an example of how CNNs are implemented
in HE using Sched-PA.
FC layers follow precisely the same steps as CNNs,
as the core primitives are also dot products.
First, the input activation ciphertext (Acts) is encoded
by placing adjacent pixels from the client's image
sequentially in polynomial slots.
This ordering eases partial ciphertext alignment.
Next, CNN filter weights (Filter) are encoded into plaintext polynomials.
Each activation-weight polynomial is multiplied with \hemult to compute the partials.
The resulting partial polynomials are then rotated to align
partial slots to the proper output-neuron slot.
Finally, with all partials computed and aligned, 
the ciphertexts are reduced with \headd.
Note how polynomial slots allow multiple output neurons to be computed
in single ciphertext.
This algorithmic parallelism provides substantial performance and memory
savings for HE as without it, each thousand degree polynomial would only compute
a \emph{single} output neuron.

The zeros found in weight plaintext slots
(e.g., PT0) ensure the correct computation.
For example, the red slot in Figure~\ref{fig:cnn_example} shows how accumulation works.
After $f_0$ is multiplied to $D_6$ in the first \hemult, 
the result is rotated right 6 times to be accumulated in the red slot ($C_{12}$).
When this rotation is performed, $D_{19}$ aligns to slot 0,
however $f_0D_{19}$ should not be accumulated in the output of slot 0 (i.e., $C_0$).
Selectively adding zeros in the plaintext slots avoids this boundary case.

\begin{figure}
\centering
\includegraphics[width=0.9\columnwidth]{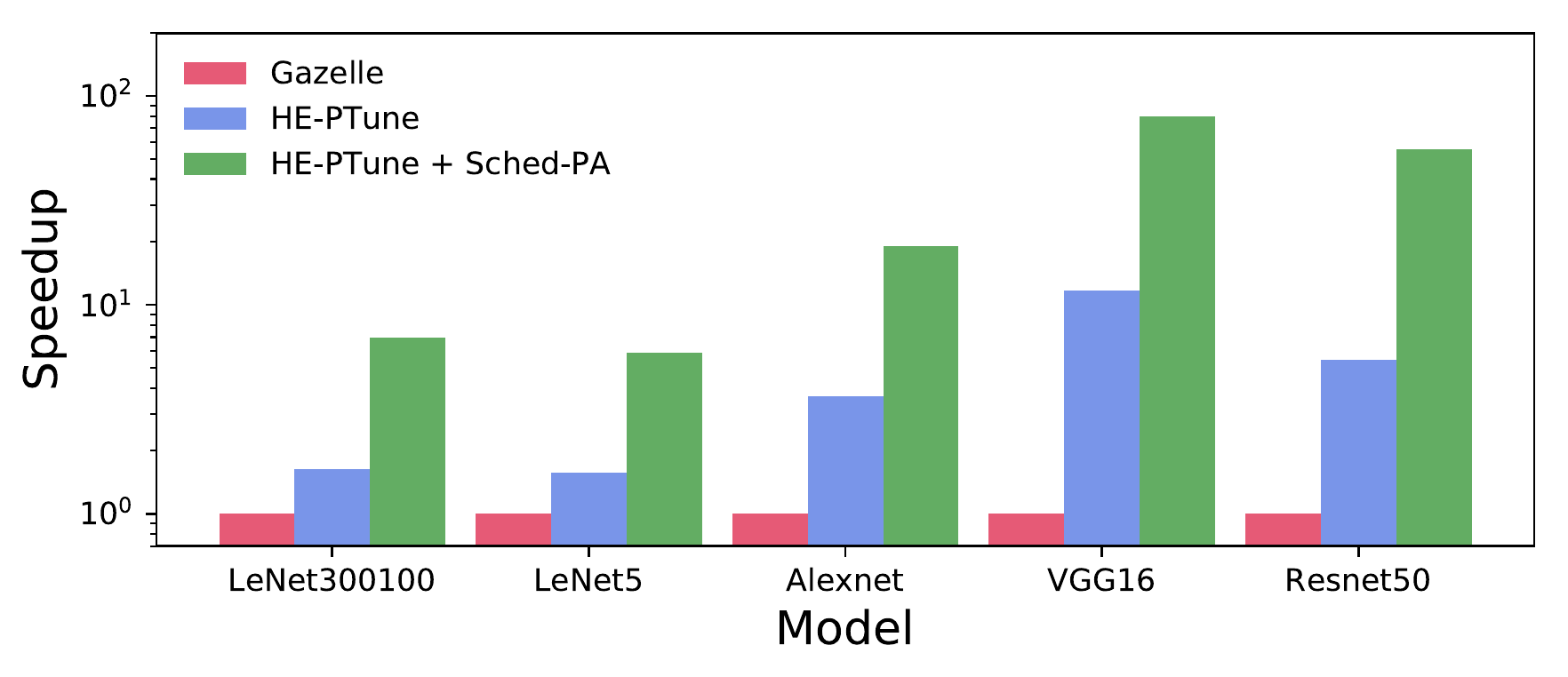}
\vspace{-1em}
\caption{
Per-benchmark speedup achieved by Cheetah using HE-PTune and Sched-PA.
Speedup is relative to Gazelle~\cite{juvekar2018gazelle}.
}
\label{fig:model_perf}
\vspace{-1.5em}
\end{figure}

\subsection{Evaluation Results}\label{sec:eval}


The effectiveness of Sched-PA is evaluated using
five standard CNN models.
HE-PTune is employed to maximize benefits
and tune HE parameters on a per-layer basis.
Multiple experiments are run to show the benefits of
HE-PTune and Sched-PA independently and relative to Gazelle.

The results for each model are shown in Figure~\ref{fig:model_perf}.
Overall, the Cheetah optimizations substantially outperform Gazelle.
Using the harmonic mean, 2.98$\times$ speedup comes from HE-PTune alone (5.25$\times$ ignoring MNIST).
Sched-PA provides an additional speedup of 5.20$\times$ (6.11$\times$ ignoring MNIST)
for a total mean performance improvement of 13.5$\times$ and maximum of 79.5$\times$ over Gazelle (30.3$\times$ mean without MNIST).

Significant performance overheads are incurred by Gazelle as
Sched-IA requires substantial ciphertext and plaintext decomposition.
Each time a polynomial is decomposed to reduce noise, 
the number of polynomials that must be computed grows proportionately.
In ResNet50, Cheetah's optimizations result 
in a ciphertext decomposition base of 8 to 16 more bits.
A higher ciphertext decomposition bases 
result in fewer decomposed polynomials for \herotate,
and substantial performance improvements.
With Sched-PA, Cheetah avoids all plaintext decomposition.


\section{Profiling HE Inference}\label{sec:profiling}

HE-PTune and Sched-PA significantly
improve the performance over the state-of-the-art~\cite{juvekar2018gazelle},
e.g., 55.6$\times$ for ResNet50.
However, with these optimizations alone HE inference is still 3-4 orders of magnitude slower than plaintext inference, i.e. unencrypted inference on a CPU.
To better understand performance bottlenecks
we profile a software implementation of HE inference
and compute the speedup needed from hardware acceleration.

We implement HE ResNet50 in the SEAL library~\cite{sealcrypto}.
Using Cheetah to tune parameters and maximize performance,
one HE inference takes 970 seconds on an Intel Xeon E5-2667 server.
The same unencrypted inference (on the same server) takes 100 milliseconds
using Keras~\cite{chollet2015keras}.
Since SEAL only supports CPUs,
we perform profiling on the CPU platform. 
Below we profile NTT running on a GPU.

\begin{figure}
\centering
\subfloat[Time breakdown]{\includegraphics[width=0.30\columnwidth]{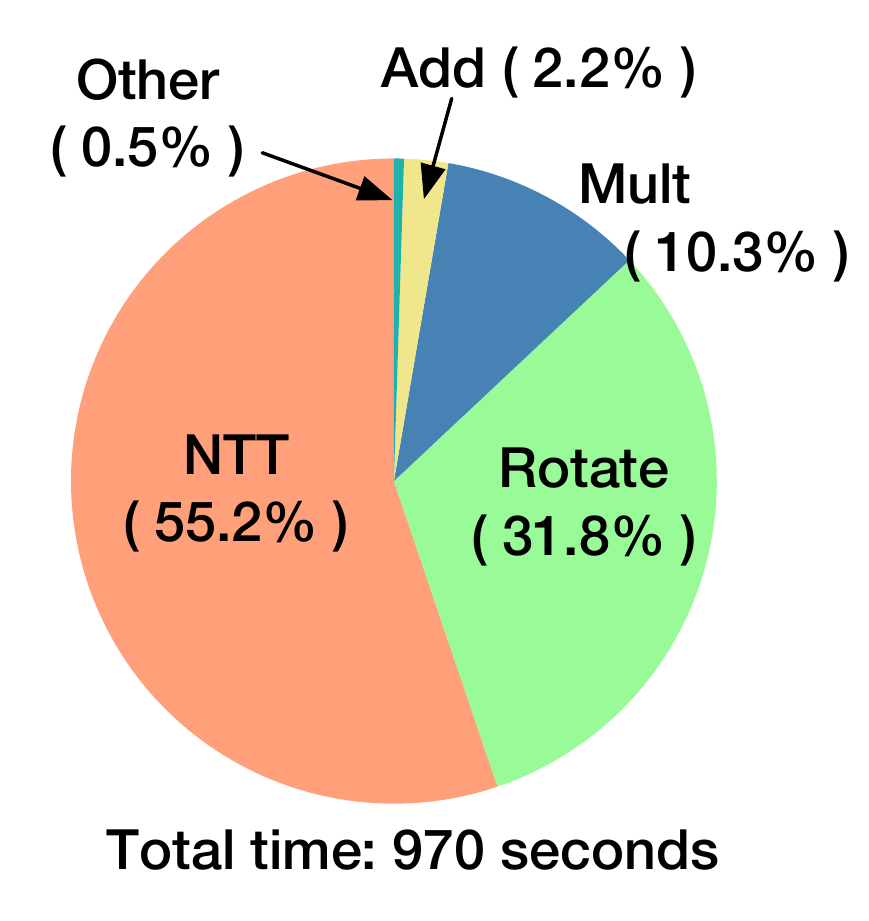}}
\subfloat[Speedup Needed]{\includegraphics[width=0.68\columnwidth]{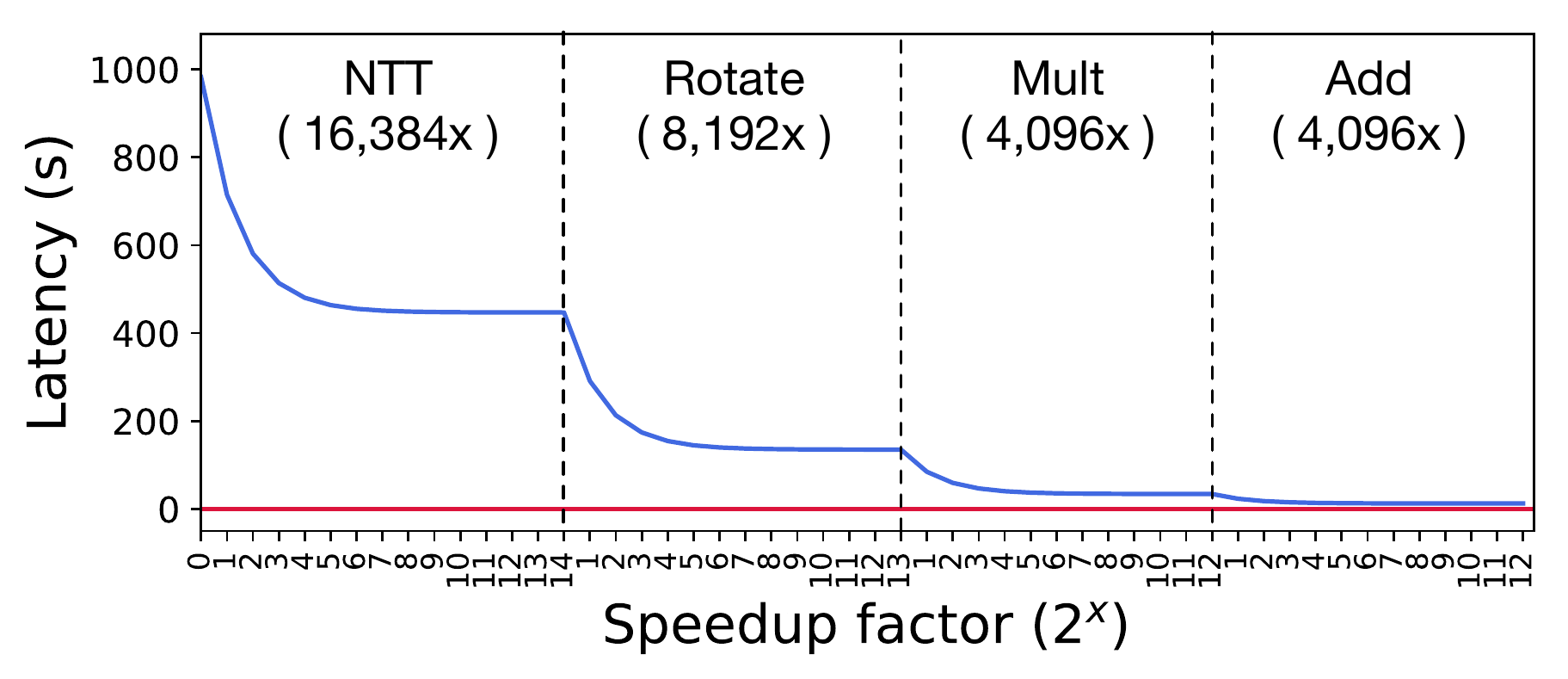}}
\vspace{-.5em}
\caption{
Profiling results for ResNet50 and speedup needed by each kernel to 
match plaintext inference latency.
}
\label{fig:profile}
\vspace{-1em}
\end{figure}

Profiling results are summarized in the pie chart of Figure~\ref{fig:profile}.
Notice that only a few kernels dominate performance
(\hemult, \headd, \herotate, and NTT).
\herotate~in Figure~\ref{fig:profile} does not include NTT as this is shown separately.
Of the four, NTT is the primary bottleneck taking 55.2\% (535 seconds) of the run time.
The SEAL profile also contains a long-tail of small functions,
labeled "Other" in Figure~\ref{fig:profile}.
We note that most of the "Other" function time is in construction/destruction.

Using the profiling results, 
we compute the speedup needed from each HE kernel to achieve plaintext inference latency.
Figure~\ref{fig:profile} shows the limit study results
and how various speedup factors impact overall run time.
The x-axis shows the speedup factor applied to each kernel function (note log scale);
the final speedup factor for each kernel is the speedup needed (e.g., 16,384 for NTT).
The y-axis shows absolute latency.
From left to right, the plot shows how the total inference latency decreases
as each theoretical speedup factor is applied to each function.
Kernel speedup is applied successively where
the run time from the most aggressive speedup factor is taken as the base for the next function.
The horizontal red line is the 100ms plaintext inference latency target.

\begin{figure}[t]
\centering
\includegraphics[width=\columnwidth]{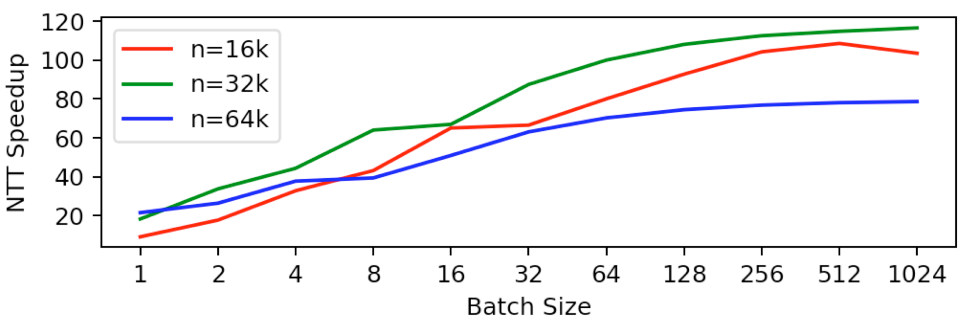}
\vspace{-2em}
\caption{{NTT GPU speedup over CPU.}} 
\label{fig:NTT_speedup}
\vspace{-2.em}
\end{figure}

\noindent \textbf{Speeding up HE with GPUs:} One way to improve kernel performance is with GPUs.
To understand the the limitations of HE on GPUs,
we benchmark NTT, the primary HE bottleneck, 
using the cuHE library~\cite{cuhe} on an NVIDIA 1080-Ti GPU.
GPU speedup is reported for different NTT batch sizes (1 to 1024) 
and vector lengths $n$ = 16K, 32K, and 64K (Figure~\ref{fig:NTT_speedup}).
At larger batch sizes (512/1024), the speedup saturates at 120$\times$.
The nvprof profiler shows that for a batch size of 512,
the GPU is utilized with 70\% warp occupancy and 85\% warp execution efficiency.

Other first order limitations to performance likely derive from
(a) non-native, long integer data types requiring emulation,
(b) modular arithmetic, which adds branch instructions and 
over 10 compute instructions per multiplication.
Despite the speedups,
GPUs fall well short of the improvements required to reach native CPU speed. 

\section{HE Inference Accelerator Architecture}
\label{sec:limits}

This section proposes a general accelerator architecture for HE inference to bridge the remaining performance gap.

{
\setlength\abovecaptionskip{3pt}
\setlength{\belowcaptionskip}{-3pt}
\begin{figure*}[t]
\begin{center}
\includegraphics[width=0.85\textwidth]{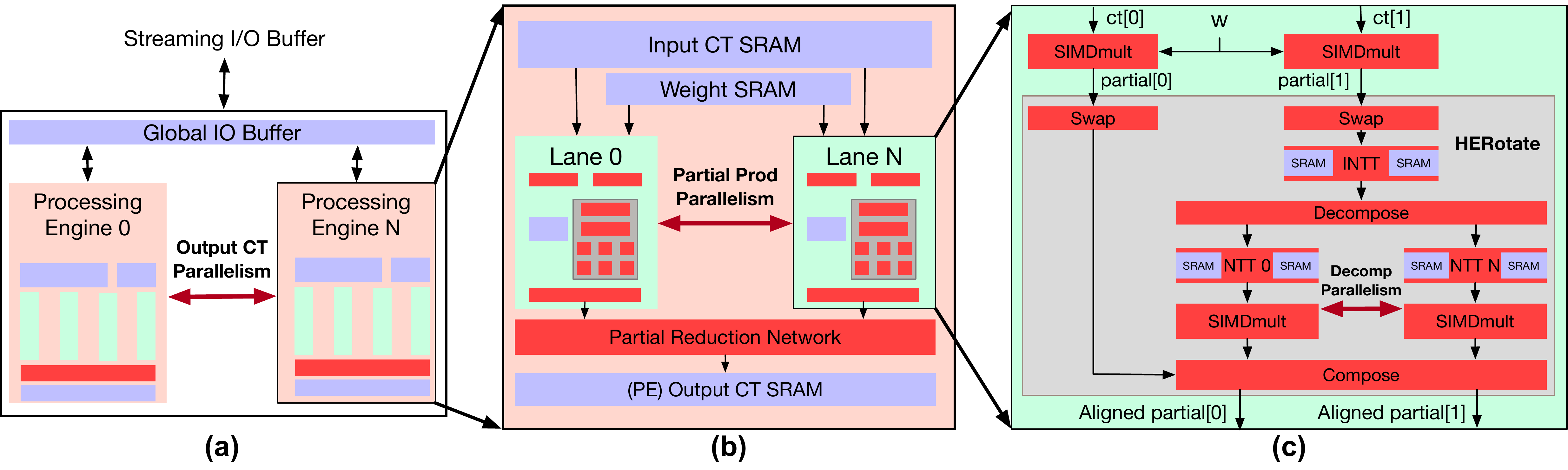}
\vspace{-.5em}
\caption{
Cheetah accelerator architecture. 
(a) The accelerator is composed of parallel PEs operating in output stationary fashion.
Off-chip data is communicated via a PCIe-like streaming interface, and data is
buffered on-chip using global PE SRAM. 
(b) Each PE contains Partial Processing Lanes which compute the HE dot product. 
(c) Lanes comprise individual HE operators.}
\label{fig:cheetah_accelerator}
\end{center}
\vspace{-2em}
\end{figure*}
}

{
\setlength{\abovecaptionskip}{3pt}
\begin{figure}[ht]
\centering
\includegraphics[width=\columnwidth]{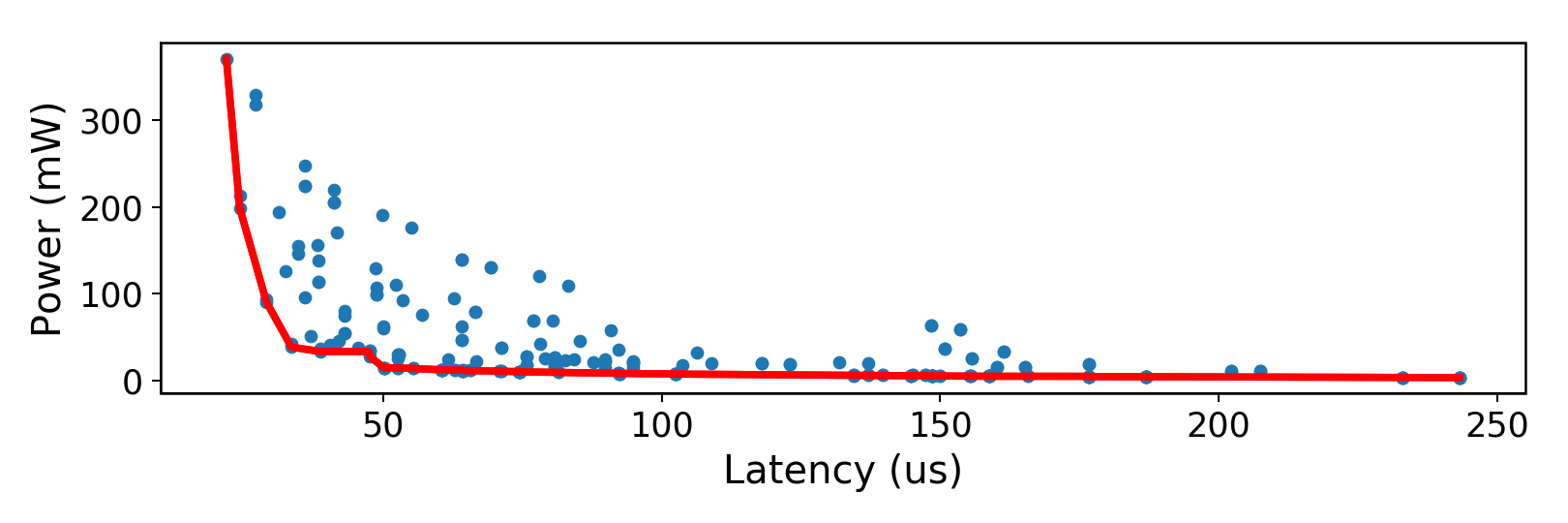}
\vspace{-1.75em}
\caption{DSE for NTT; Pareto frontier in red.}
\label{fig:NTT_pareto}
\vspace{-2em}
\end{figure}
}

\subsection{Accelerator Architecture}

The proposed accelerator architecture is shown in Figure~\ref{fig:cheetah_accelerator}.
At a high level, it is composed of ciphertext (CT) \emph{processing engines} (PEs)
that receive data from a PCIe-like streaming interface and buffer intermediary results in SRAMs (Figure~\ref{fig:cheetah_accelerator}a).
Hierarchically, PEs are composed of \emph{partial processing lanes} or \emph{Lanes}, and a \emph{partial reduction network}, which implement the HE dot product (Figure~\ref{fig:cheetah_accelerator}b).
Lanes are further decomposed into individual HE operators (Figure~\ref{fig:cheetah_accelerator}c).

\subsubsection{PEs: Output Neuron Engines}

Our architecture is designed to maximize performance and parallelism by being output-stationary.
Each PE processes a single ciphertext of output neurons
and all compute-memory resources necessary for the output remain local to the PE; the number of PEs is parameterized.
When there are more output neurons per layer (Parallel Output CTs)
than physical PEs, we time multiplex the computation across multiple PE executions.
The PEs are connected to input and output buffers
used to route data to and from the host.
These buffers constitute small SRAMs as they only handle communication, 
all state and intermediates are local to PEs.

The internals of the PE contain partial processing lanes and reduction networks.
Each PE contains an \emph{Input CT buffer} to store a copy of all activation CTs locally, a relatively small SRAM for weights, a set of partial processing lanes, 
a partial reduction network, and output CT SRAM.
Each Lane is capable of processing a unique dot product partial; 
the number of lanes is parameterizable.
Lanes within a PE operate in lockstep to enable aggressive reuse of twiddle factor SRAMs required for NTTs.
The partial reduction network is configured based on the number of partials
computed in parallel (i.e., number of Lanes). 
Input CT SRAMs are provisioned with enough capacity to hold all the inputs
with sufficient bandwidth to feed all Lanes.

\subsubsection{Lanes: Partial Engines}

Lanes are the backbone of the accelerator and implement the HE operators.
In Figure~\ref{fig:cheetah_accelerator}c, HE kernel blocks are denoted in red.
Intermediary SRAMs, shown in blue, are used to store results between HE kernels.
We use SRAMs instead of off-chip DRAM for intermediary results because of the high internal bandwidth required within NTT modules to support aggressive parallelism. 
In the worst case, each NTT kernel requires 13 GB/s of combined internal bandwidth; each lane contains multiple NTTs and each PE contains many lanes.
Aside from the NTT kernels that require a strided access pattern, 
all operations within a Lane can be made streaming (i.e, no SRAMs needed after kernels).
This allows the architecture to save SRAM resources.
The NTT activation decomposition factor $A_{dcmp}$ introduces a parametrizable degree of inter-NTT parallelism, which otherwise does not impact overall latency.
For high-performance, enough lanes are allocated to
execute all partials in parallel.


The lane architecture shows the datapath and dependencies 
to compute a single partial dot product.
Both input polynomials (CT[0] and CT[1]) are first multiplied by plaintext weights using the \hemult~ operator, outputting partial polynomials.
The datapaths diverge as BFV splits the compute asymmetrically 
between partial[0] and partial[1].
\herotate~ is applied to perform polynomial slot alignment.
For partial[1], inverse NTT (INTT), decomposition, NTT, and composition units are applied.
The datapath for partial[1] splits after the INTT computation in order to implement ciphertext decomposition.
Recall that decomposition reduces noise growth;
however, the trade-off manifests here as additional compute requirements.
Fortunately, the additional computation can be parallelized (NTTs and SIMDmults).
The decomposed polynomials are then converted back to the evaluation domain and combined with swapped partial[0] to produce the aligned partial that is fed to the partial reduction network in the PE, which consists of SIMDadd units.
\section{Experimental Results}
\label{sec:results}

This section presents design space exploration results of the parameterized accelerator architecture.
We show that by combining Cheetah's algorithmic and architectural optimizations
provides near plaintext performance.

{
\setlength{\belowcaptionskip}{-2pt}
\begin{figure*}
\centering
\subfloat[ResNet50 DSE]{\includegraphics[width=0.33\textwidth]{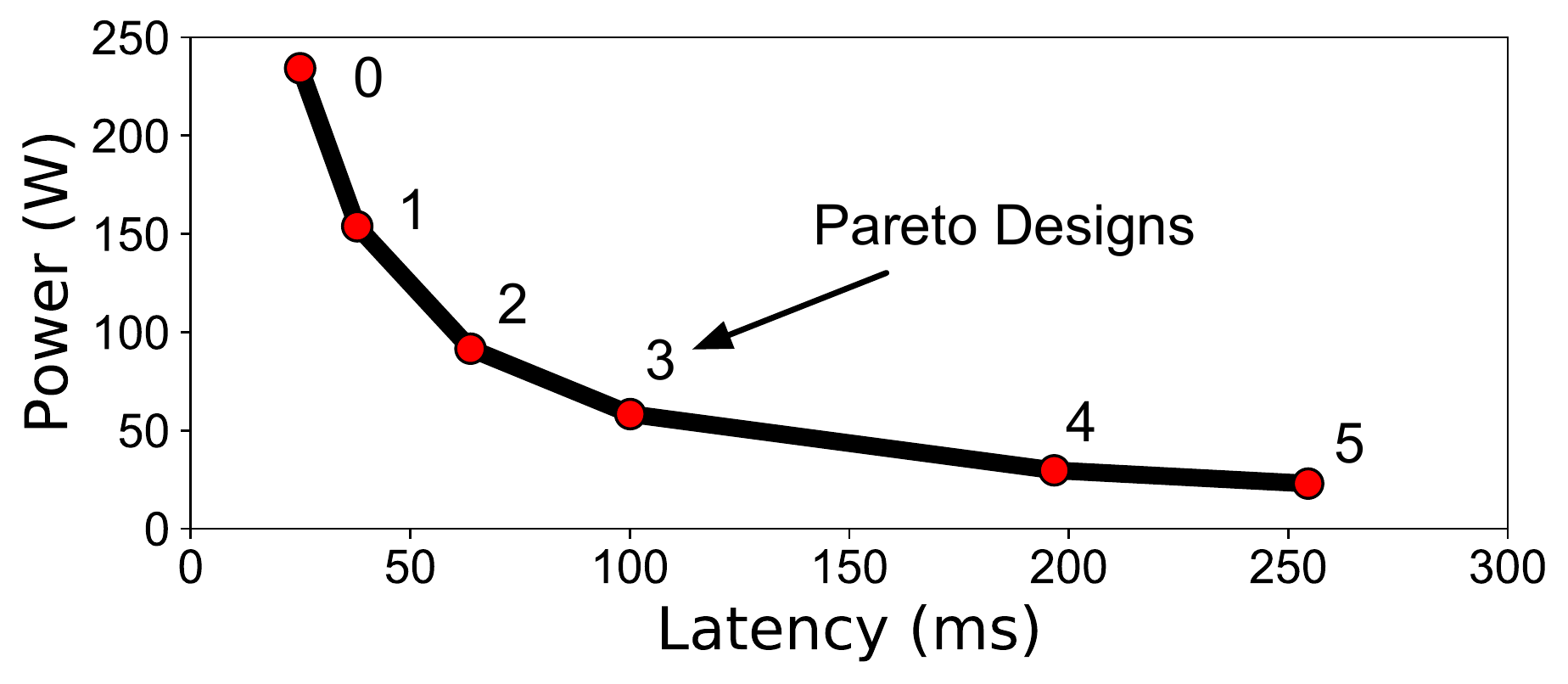}}
\subfloat[Time Breakdown]{\includegraphics[width=0.33\textwidth]{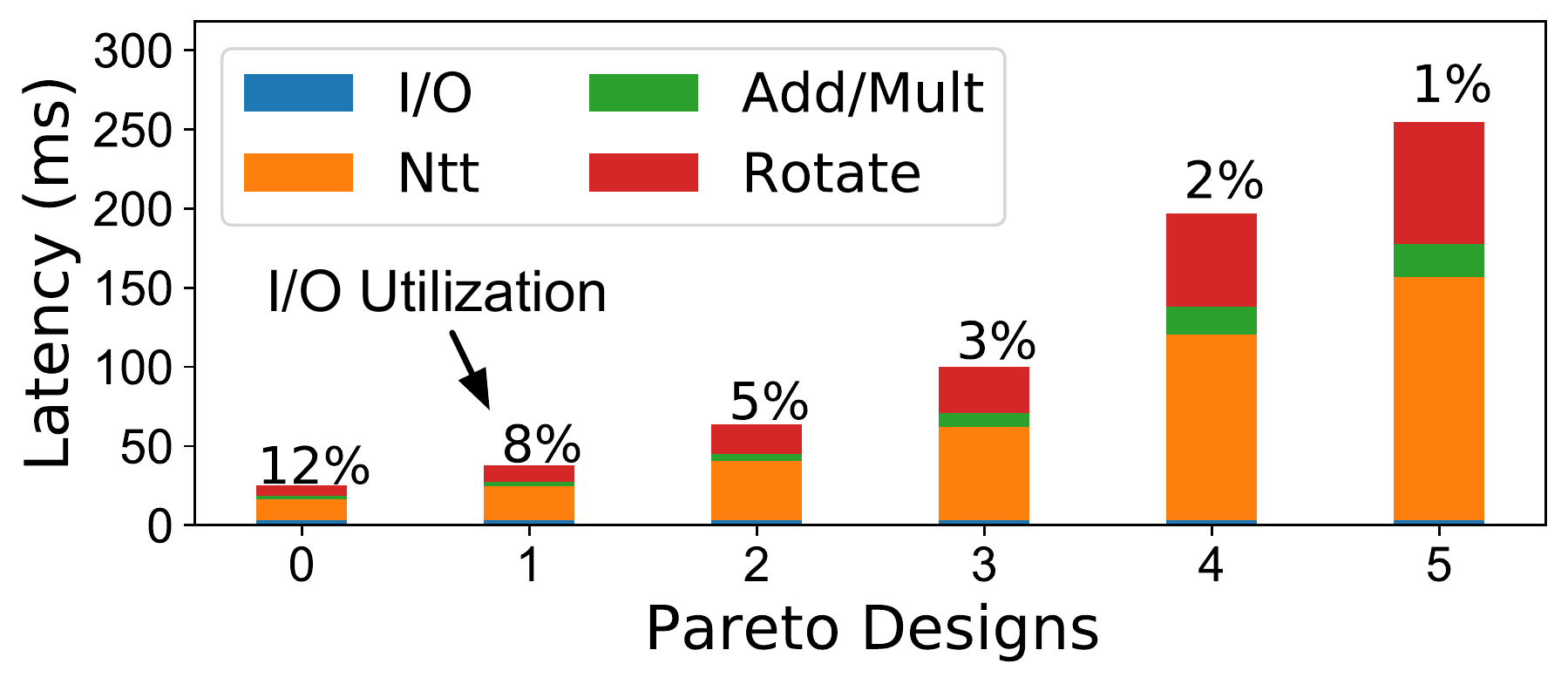}}
\subfloat[Area Breakdown]{\includegraphics[width=0.33\textwidth]{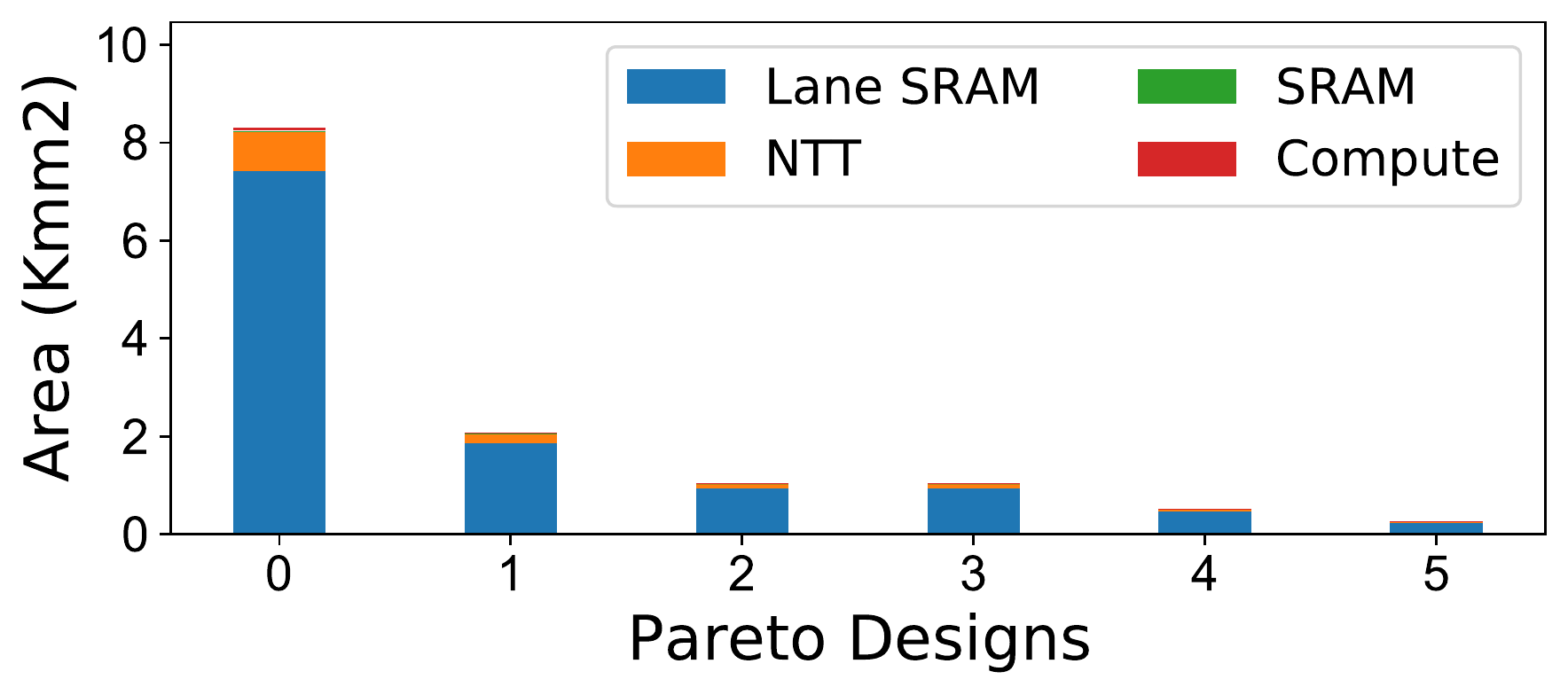}}
\vspace{-0.5em}
\caption{
(a) Power-latency Pareto for ResNet50 DSE. (b) Run time breakdown for each Pareto design point. (c) Energy and area breakdown for each Pareto design point.
}
\label{fig:acc_dse}
\vspace{-1.5em}
\end{figure*}
}

\subsection{Methodology}

Design space exploration consists of sweeping
various accelerator microarchitectural parameters.
Each kernel 
(\hemult, \headd, and \herotate--which is split into Swap, INTT, Decompose, NTT, SIMDMult, and Compose)
is built using Catapult HLS 10.3d~\cite{catapult_hls} 
and synthesized with a commercially-available 40nm standard cell library targeting 400 MHz. 
For each kernel, we evaluate hundreds of design points
to explore different design tradeoffs and identify optimal implementations.
Each kernel accelerator's microarchitecture 
is parameterized by memory bandwidth (or IO in the case of streaming kernels), 
datapath parallelism (i.e., hardware loop unrolling),
pipelining (i.e., initiation-interval), 
and clock frequency.
We estimate power, performance, and area
using Catapult's output RTL and power analysis flows.
We use a commercial SRAM compiler to compile each SRAM dimension used across different design points due to different memory tiling factors.

Based on these kernel design sweeps, we select Pareto optimal points and use them to further identify optimal HE accelerator designs with a simulator.
The simulator models the computations and hardware described in Section~\ref{sec:limits}.
It takes as input HE parameter settings from HE-PTune and
user defined accelerator microarchitectural parameters (HE kernel implementation, number of PEs, and number of lanes).
We swept PEs per accelerator are swept from 2-1024 and
lanes per PE from 4-8192.
Area is estimated based on architectural parameters while power and performance are derived through simulation.
To estimate performance and power for an input DNN, 
each layer is represented as the number of input/output ciphertexts and partials per output ciphertext.
The simulator then maps and multiplexes the number of output neuron ciphertext
to available PEs and partials to lanes to derive hardware activity factors.
Combining multiplexing and activity factors with HLS latency and power results estimates accelerator performance.

The overall performance of a full inference is modeled on a per-layer granularity; this is because after each layer's linear computations, activations are sent to the client for ReLU and Pooling.
Each layer of a DNN is expressed as a series of output neuron computations; from this, we compute the total number of partials per output neuron ciphertext.


To capture the benefits of technology scaling, 
we report power and area estimates for 5nm using foundry-reported scaling factors.
Specifically, we use 0.2$\times$ power and 0.22$\times$ area to scale from 40nm to 16nm,
based on~\cite{TSMCscalingIEEE, tsmcnode28nm, tsmcnode20nm, tsmcnode16nm}. 
From 16nm to 5nm, the power and area scaling factors are 0.32$\times$ and 0.17$\times$, using~\cite{tsmcnode5nm} and recent data from~\cite{geoffrey2019tsmc}.
Together, the power and area scaling factors (40nm to 5nm) are 0.056$\times$ and 0.038$\times$, respectively.

\subsection{Evaluation Results}


\subsubsection{Intra-Kernel Parallelism}

We measure the Pareto optimal design speed ups for HE kernels achieved by the energy-optimal point from the power-latency Pareto frontier.
An example design space Pareto frontier for NTT is shown in Figure{~\ref{fig:NTT_pareto}}.
Recall that these frontiers are used as the cost model for the larger architecture, whose sweeps consider the performance-latency tradeoffs of each kernel.
We normalize our speedups to the SEAL library implementation on a 3GHz Intel SkyLake Xeon processor.

We see modest speedups of individual kernels up to 40$\times$ speedup (average $\approx$10$\times$ speedup) with hardware acceleration.
The {\headd} and {\hemult} kernels provide substantial parallelism as the underlying computation
consists of element-wise modular additions and multiplications which are easily parallelized.
In the case of {\herotate~}, (Swap, Decompose, Compose) and NTT, 
parts of the underlying computations occur sequentially 
while others can be parallelized such as the element-wise multiplications and butterfly computations. 
The key result is that \emph{intra-kernel parallelism can reduce HE overhead by roughly one order of magnitude}.



\subsubsection{Inter-Kernel Parallelism}
Fortunately, DNNs contain abundant parallelism.
With the exception of kernel dependencies within a Lane and the reduction
of partial products in PEs, partials and output neurons can be executed in parallel by allocating more hardware resources.

For example, consider CNN Layer6 in ResNet50 ($f_w=3$, $w=64$, $c_i=c_o=64$). 
If each ciphertext contains a single input channel ($n=4096$), then all partial products can be computed with 36,864 \hemult~and \herotate~parallel kernel invocations.
The partial products for these layers cannot be parallelized since \hemult~must be performed before \herotate~under Sched-PA.
In \herotate~, domain conversion from evaluation to coefficient using INTT must be done before decomposition, but the NTT to convert back the domain of decomposed polynomials can be parallelized.
As a result, we find that \emph{the degree of parallelism that can be exploited at the Lane and PE
level is on the order of thousands for ResNet50}.
The key result is that application \emph{inter-kernel parallelism exposes two to three order of magnitude improvement}.

\subsubsection{Lane and PE DSE}

When combined, \emph{inter-kernel and intra-kernel parallelism can bridge the remaining 3-4 order of magnitude speedup required to approach plaintext inference speeds on top of HE-PTune and Sched-PA}.
We conduct a design space exploration to evaluate whether these designs are practical with respect to power and area.
Figure~\ref{fig:acc_dse} shows the results from the design space exploration of ResNet50.
The power-latency Pareto points identified in the left-most subplots are the ideal architectures when designing an accelerator tuned only for the model.
The Pareto frontiers provide insight into the hardware cost-per-ms tradeoff of inference latency.
For ResNet50, we find the Pareto optimal design point requires around 30W and 545mm$^2$ for ResNet50, 
which are within feasible (albeit high) resource usage of datacenter coprocessors.
The low power density is due to aggressive SRAM tiling to meet aggressively high internal bandwidth targets for NTT units.
Upon further analysis, we find that the 128$\times$60 bit SRAM sizes have a bit density that is $\approx2.5\times$ worse than larger 1024$\times$60 SRAMs, which results in low power density.
We also note that the 400 MHz clock target is low for a 5nm technology, furthering reducing power density.

To understand the limitations to efficiency and performance of each Pareto design point, Figure~\ref{fig:acc_dse} shows the Pareto optimal design result for ResNet50 (AlexNet, VGG16, and MNIST exhibit similar trends).
Figure~\ref{fig:acc_dse}a shows six design points on the Pareto frontier.
Figure~\ref{fig:acc_dse}b and Figure~\ref{fig:acc_dse}c show the breakdown of run time and area respectively for these six design points.
For extreme low-latency designs (Pareto points 0 and 1), results show that most of the design area goes into small SRAMs which are required to support the enormous internal bandwidth required by NTT units (discussed next).
As a result, this leads to impractically large area overheads.

Overall, the results in Figure~\ref{fig:acc_dse}b confirm
NTT and reduction (\herotate) dominate HE accelerator computation cost.
Recall NTT is data intensive and have many small internal SRAMs,
which at extreme design points result in high power and area usage.
This is compounded by the sheer number of NTT units that operate in parallel, 
making NTT computations the largest overall area component. 
We note that even in the most parallel design point considered, 
the accelerator is compute bound (IO utilization is only 12\%)
and NTT remains the primary bottleneck.
Moreover, we find that the input and output SRAMs in the architecture do not incur as high of a power and area cost. 
This means that the input duplication into each PE to support output-stationary computation is relatively inexpensive.

\begin{table}[t]
\begin{center}
\begin{small}
\caption{Performance of running VGG16 and AlexNet
on PT-ResNet50 accelerator. Prt is partials per output CT.}
\label{tab:flex_study}
\vspace{-0.5em}
\resizebox{\columnwidth}{!}{\begin{tabular}{c c c c c c c}
\toprule
Model    & Lat(ms) & Increase  & PEs-Lanes  & Out CT $\mu $ & (K)Prt $\mu $ \\
\midrule
ResNet50 & 100     &    0\%    &   8-512    & 147            &  50.5        \\
VGG16    & 215     &   59\%    &  16-256    & 422            &  595          \\
AlexNet  &  77     &   28\%    &  16-128    & 475            &  337          \\
\bottomrule
\end{tabular}}
\end{small}
\end{center}
\vspace{-1em}
\end{table}

\subsubsection{Accelerator Generality}

Designing a fixed-size accelerator for each DNN model is impractical.
Instead, the accelerator can be programmed to support different-sized networks by multiplexing compute logic (PEs and Lanes) to handle different DNN tensor shapes.
To quantify the loss associated with underutilized units and imperfect
dimension matching, we measure performance loss for different ImageNet
models (AlexNet and VGG16) running on the HE accelerator optimized for ResNet50 from Figure~\ref{fig:acc_dse} (i.e., Point 3).

The performance results are summarized in Table~\ref{tab:flex_study}.
We find both AlexNet and VGG16 experience considerable slowdown
relative to their ideal architectures for real-time inference as seen in the {\it Increase} column of the table.
This is due to the choice of PE and Lane allocations and the differences in layer dimensions.
As the table shows, AlexNet and VGG16 layers have a higher average number of output CTs per layer than ResNet50, and the cost of multiplexing PEs 
outweighs the cost of poor Lane utilization given the granularity of work 
(a partial versus an entire output CT).
The average number of partials per ciphertext is also much higher. 
However, ResNet50 is very structured given its use of bottleneck layers
many of which have partials per output ciphertext that are divisible by or less than 512, yielding high utilization.
Conversely, VGG16's partials per output CT 
tend to fall just above or below factors of 512,
(e.g., 34, 687, 1086) resulting in lower utilization.

\section{Related Work}
\label{sec:related_work}

A growing interest in privacy and machine learning has
resulted in a body of related work on developing cryptographic solutions.
Techniques can be categorized into two groups: 
HE only~\cite{gilad2016cryptonets,hesamifard2017cryptodl,sanyal2018tapas,brutzkus2019low},
or multiparty computation (MPC)-based~\cite{rouhani2017deepsecure,liu2017oblivious,riazi2018chameleon,juvekar2018gazelle}.
While each has significantly advanced the field
all suffer from either accuracy loss due approximation
or high communication/computation overheads.

\textit{HE only} techniques must address evaluating non-linear functions 
(e.g., ReLU, MaxPool) using only available addition and multiply operations.
CryptoNets~\cite{gilad2016cryptonets}, CryptoDL~\cite{hesamifard2017cryptodl}, and LoLa~\cite{brutzkus2019low} 
propose replacing ReLU with low-order polynomials that can readily be computed with HE primitives.
However, even with square activations~\cite{brutzkus2019low},
this requires very large HE parameters 
(e.g., $q$ $\approx$ 1000~\cite{gilad2016cryptonets}, 440 bits~\cite{brutzkus2019low}, while Cheetah uses 60)
for an appropriate noise budget.
Moreover, approximate activation functions require re-training~\cite{brutzkus2019low}
and can degrade accuracy~\cite{gilad2016cryptonets}.
Others propose accelerating HE kernels with accelerators.
NTT has been ported to FPGAs~\cite{poppelmann2015accelerating,roy2019fpga} 
and GPUs~\cite{akleylek2015efficiency,al2018hpfv,dai2015cuhe} to speedup polynomial multiplication.
Raizi et al~\cite{heax} propose HEAX to accelerate HE kernels with FPGAs but only reports two orders of magnitude speedup.
While related, the results of HEAX are orthogonal to the contributions of this paper; 
HEAX uses CKKS (Cheetah uses BFV). Mostly, above works
focus on ciphertext-ciphertext multiplication (Cheetah uses ciphertext-plaintext), and targets kernel acceleration 
(Cheetah focus on the application of DNN inference and general chip architecture).

\textit{MPC-based} schemes provide an alternative to approximation by combining HE with
other security solutions, typically garbled circuit (GC)
\cite{orlandi2007oblivious,7958569,rouhani2017deepsecure,liu2017oblivious,riazi2018chameleon,juvekar2018gazelle}.
Among them, Gazelle is considered the state-of-the-art~\cite{juvekar2018gazelle}.
Gazelle uses HE for linear layers in the cloud and 
GC~\cite{yao1986generate} for ReLU and MaxPool on the client.
This can significantly improve the latency for small models
but results in a severe computational bottleneck in deep models (e.g., ResNet50).
Cheetah takes Gazelle as a baseline and
focuses on reducing the significant computational overheads of HE.

Other work assumes different threat models with non-cryptographic solutions. 
E.g.,~\cite{tramer2018slalom,brasser2018voiceguard} use
TEEs to isolate private data from untrusted software.
Others have looked at limiting information leakage by adding noise (similar to DP)~\cite{mireshghallah2019shredder};
this provides increased average-case privacy with negligible loss in accuracy.
\section{Conclusion}
\label{conclusion}

This paper makes progress on the fundamental roadblocks to adoption 
of HE machine learning inference by proposing Cheetah. 
First, 
HE-PTune automatically identifies 
the HE parameter with minimal computational overhead. 
By fine-grained tuning of HE parameters per each layer, 
HE-PTune’s parameters yield up to 11.7$\times$ performance benefit over the state-of-the-art.
Next, Cheetah proposes Sched-PA which yields 10.2$\times$ speedup for dot products in FC and CNN layers in HE.
Finally, we propose a custom accelerator architecture to exploit the high application inter- and intra-kernel parallelism. 
We evaluate the tradeoffs between inference latency and hardware costs and show that combining application level parameter tuning with specialized hardware acceleration can bring HE inference down to practical inference speeds.




\bibliographystyle{IEEEtranS}
\bibliography{refs}

\end{document}